\documentclass[aps,prd,preprint,floats,epsf,superscriptaddress,nofootinbib]{revtex4}
\pdfoutput=1
\usepackage{textcomp}
\usepackage{amssymb}
\usepackage{anysize}
\usepackage{bold-extra}
\usepackage{enumerate}
\usepackage{graphicx}
\usepackage{lscape}
\usepackage{mathrsfs}
\usepackage{multirow}
\usepackage{hhline}
\usepackage{makecell}
\usepackage[final]{pdfpages}
\usepackage{rotating}
\usepackage{subfig}
\usepackage{url}
\usepackage{wrapfig}
\usepackage{amsmath}
\usepackage{color}
\usepackage{fancyhdr}
\usepackage{verbatim}
\usepackage{epstopdf}
\usepackage[section]{placeins}
\usepackage[normalem]{ulem} 
\usepackage[colorlinks=true,
            linkcolor=red,
            urlcolor=blue,
            citecolor=blue]{hyperref}
\def\nn{\nonumber}
\def\bea{\begin{eqnarray}}
\def\eea{\end{eqnarray}}
\def\ba{\begin{eqnarray}}
\def\ea{\end{eqnarray}}
\def\be{\begin{equation}}
\def\ee{\end{equation}}
\def\beq{\begin{equation}}
\def\eeq{\end{equation}}

\newcommand{\tf}{\texorpdfstring}

\newcommand{\gev}{~\text{GeV}}

\newcommand{\fb}{~\text{fb}}

\newcommand{\abi}{~\text{ab}^{-1}}
\newcommand{\fbi}{~\text{fb}^{-1}}

\def \eewwh{$e^+e^-\to W^+W^-H$~}
\def \llljj{$\ell^{\pm}\ell^{\pm}\ell^{\mp}jj$~}
\def \jjlbb{$jj\ell^{\pm}bb$~}

\begin{document}

\title{Measuring the ratio of $HWW$ and $HZZ$ couplings \\
through $W^+W^-H$ production}

\author{Cheng-Wei Chiang}
\email[e-mail: ]{chengwei@phys.ntu.edu.tw}
\affiliation{Department of Physics, National Taiwan University, Taipei, Taiwan 10617, R.O.C.}
\affiliation{Institute of Physics, Academia Sinica, Taipei, Taiwan 11529, R.O.C.}
\affiliation{Physics Division, National Center for Theoretical Sciences, Hsinchu, Taiwan 30013, R.O.C.}

\author{Xiao-Gang He}
\email[e-mail: ]{hexg@phys.ntu.edu.tw}
\affiliation{Department of Physics, National Taiwan University, Taipei, Taiwan 10617, R.O.C.}
\affiliation{Physics Division, National Center for Theoretical Sciences, Hsinchu, Taiwan 30013, R.O.C.}
\affiliation{T.-D. Lee Institute and INPAC, School of Physics and Astronomy, Shanghai Jiao Tong University, Shanghai 200240, China}

\author{Gang Li}
\email[e-mail: ]{gangli@phys.ntu.edu.tw}
\affiliation{Department of Physics, National Taiwan University, Taipei, Taiwan 10617, R.O.C.}

\date{\today}

\begin{abstract}
For a generic Higgs boson, measuring the relative sign and magnitude of its couplings with the $W$ and $Z$ bosons  is essential in determining its origin.  Such a test is also indispensable for the 125-GeV Higgs boson.  We propose that the ratio of the $HWW$ and $HZZ$ couplings $\lambda_{WZ}$ can be directly determined through the $W^+W^-H$ production, where $H$ denotes a generic Higgs boson, owing to the tree-level interference effect.  While this is impractical at the LHC due to the limited sensitivity, it can be done at future $e^+e^-$ colliders, such as a 500-GeV ILC with the beam polarization $P(e^-,e^+)=(-0.8,+0.3)$ in the \jjlbb and \llljj channels.  The discovery potential of a general ratio and the power to discriminate it from the SM value are studied in detail. Combining the cross section of \eewwh with the measurements of $HZZ$ coupling at the HL-LHC, one can further improve the sensitivity of $\lambda_{WZ}$.

\end{abstract}

\maketitle



\section{Introduction}

After the discovery of the 125-GeV Higgs boson~\cite{Aad:2012tfa,Chatrchyan:2012xdj}, it is important to measure all its couplings with other particles to verify whether it is the Standard Model (SM) Higgs boson.  In particular, the measurements of $HVV$ couplings are crucial to testing exactly how electroweak symmetry breaking (EWSB) occurs.  In the SM, there is a residual global $SU(2)$ symmetry known as the custodial symmetry~\cite{Sikivie:1980hm} in the Higgs and gauge sectors after the EWSB to guarantee that the electroweak $\rho$ parameter
\begin{align}
\rho \equiv \frac{m_W^2}{m_Z^2\cos^2\theta_W} = 1
\end{align}
at tree level, where $\theta_W$ is the weak mixing angle, and $m_Z$ and $m_W$ are the $Z$ and $W$ boson masses, respectively.  It is also found that the the tree-level $HWW$ and $HZZ$ couplings satisfy~\cite{Willenbrock:2004hu} 
\begin{align}
\label{eq:sm_ratio}
\dfrac{g_{HWW}^{\text{SM}}}{g_{HZZ}^{\text{SM}}\cos^2\theta_W}=1
\end{align}
as $g_{HWW}^{\text{SM}}=2m_W^2/v^2$, $g_{HZZ}^{\text{SM}}=2m_Z^2/v^2$, where $v = 246$~GeV is the vacuum expectation value (VEV) of the Higgs field.

For a neutral Higgs boson $H$ that may have an origin beyond the SM, we write its couplings with the $W$ and $Z$ bosons as
\begin{align}
g_{HWW}=\kappa_W g_{hWW}^{\text{SM}},\quad g_{HZZ}=\kappa_Z g_{hZZ}^{\text{SM}},
\end{align}
where the coefficients $\kappa_W$ and $\kappa_Z$ are the scale factors~\cite{Heinemeyer:2013tqa,Mariotti:2016owy}.  Note that $H$ can generally be any neutral Higgs boson, though our following discussions will focus on the 125-GeV Higgs boson for definiteness and obvious interest.  If $H$ is the SM Higgs boson, then $\kappa_W=\kappa_Z=1$.  However, $\kappa_{W,Z}$ can have generally different values for $H$ with origin from an extended Higgs sector.  It is therefore useful to consider the ratio
\begin{align}
\lambda_{WZ}\equiv \kappa_W/\kappa_Z 
~.
\end{align}

In Higgs-extended models, even if $\rho=1$ is imposed at tree level, the ratio $\lambda_{WZ}$ can be different from 1 and even have a negative sign if at least two distinct non-trivial Higgs representations are introduced~\cite{Gunion:1990kf,Chen:2016ofc,Cen:2018wye}.  Ref.~\cite{Cen:2018wye} has shown that $\lambda_{WZ}$ is generally arbitrary when the Higgs sector of the SM is extended with a complex triplet and a real triplet while keeping $\rho=1$. It is therefore important to be able to determine the magnitude and sign of $\lambda_{WZ}$ for any neutral Higgs boson, including the 125-GeV Higgs boson.

Experimentally, it is difficult to directly determine the sign of $\lambda_{WZ}$.  Existing measurements~\cite{Khachatryan:2016vau,CMS-PAS-HIG-17-031} of the $HWW$ and $HZZ$ couplings, depending on the production or decay rates, are more sensitive to the magnitude rather than the sign of $\lambda_{WZ}$.  From the combined measurements in Large Hadron Collider (LHC) Run I~\cite{Khachatryan:2016vau}, the $2\sigma$ confidence level (C.L.) interval of $\lambda_{WZ}$ is determined to be
\begin{align}
-1.10\lesssim \lambda_{WZ} \lesssim -0.73\quad \text{or} \quad 0.72\lesssim \lambda_{WZ} \lesssim 1.10,
\nonumber
\end{align}
while the result from LHC Run II with the integrated luminosity of $35.9\fbi$ is~\cite{CMS-PAS-HIG-17-031,ATLAS-CONF-2018-002,ATLAS-CONF-2018-004}
\begin{align}
-1.39\lesssim \lambda_{WZ} \lesssim -0.97\quad \text{or} \quad 0.92\lesssim \lambda_{WZ} \lesssim 1.37.
\nonumber
\end{align}
It is clear that there is a discrete two-fold ambiguity in the sign of $\lambda_{WZ}$ that cannot be resolved in such experiments.

At the High-Luminosity LHC (HL-LHC) with an integrated luminosity of $3\abi$, the $HWW$ and $HZZ$ couplings are expected to be measured with the precision~\cite{CMS:2013xfa}
\begin{align}
\delta\kappa_W/\kappa_W \leq 5\%,\quad \delta\kappa_Z/\kappa_Z \leq 4\%,
\end{align}
assuming that the central values are the ones predicted by the SM~\cite{Peskin:2013xra}.
The corresponding $1\sigma$ uncertainty on $\lambda_{WZ}$ is then
\begin{align}
\label{eq:precision_lam}
{\delta\lambda_{WZ}}/{\lambda_{WZ}} \leq 6.4\%
~.
\end{align}

It was investigated in Ref.~\cite{Chen:2016ofc} that the magnitude and sign of $\lambda_{WZ}$ could be measured in the differential distribution of $H\to ZZ^{*}\to 4\ell$ due to the interference between amplitudes at tree and one-loop levels, which are proportional to the $HZZ$ and $HWW$ couplings, respectively.  However, more diagrams should be considered at the one-loop level~\cite{Kniehl:1990mq,Bredenstein:2006ha} to maintain gauge invariance and consistency~\cite{Mariotti:2016owy}.  In this work, we show that a desirable interference occurs among tree-level amplitudes in the $W^+W^-H$ production process.  We therefore propose to use this channel to experimentally fix $\lambda_{WZ}$.

This paper is arranged as follows. In Section~\ref{sec:constraint}, we discuss the current constraints on the ratio $\lambda_{WZ}$ of the $HWW$ and $HZZ$ couplings and the projected sensitivity at the HL-LHC.  We argue why it is impractical to employ the proposed method at the LHC.  In Section~\ref{sec:ILC}, we study $\lambda_{WZ}$ in the \eewwh process at a future $e^+e^-$ collider with the colliding energy of 500~GeV.  We separately consider two kinds of final states, \jjlbb and \llljj, and study their reaches.  Section~\ref{sec:summary} summarizes our results.

\section{\tf{$W^+W^-H$}{WWH} production at hadron colliders}
\label{sec:constraint}

\begin{figure}[!htb]
\centering
\includegraphics[width=0.8\textwidth]{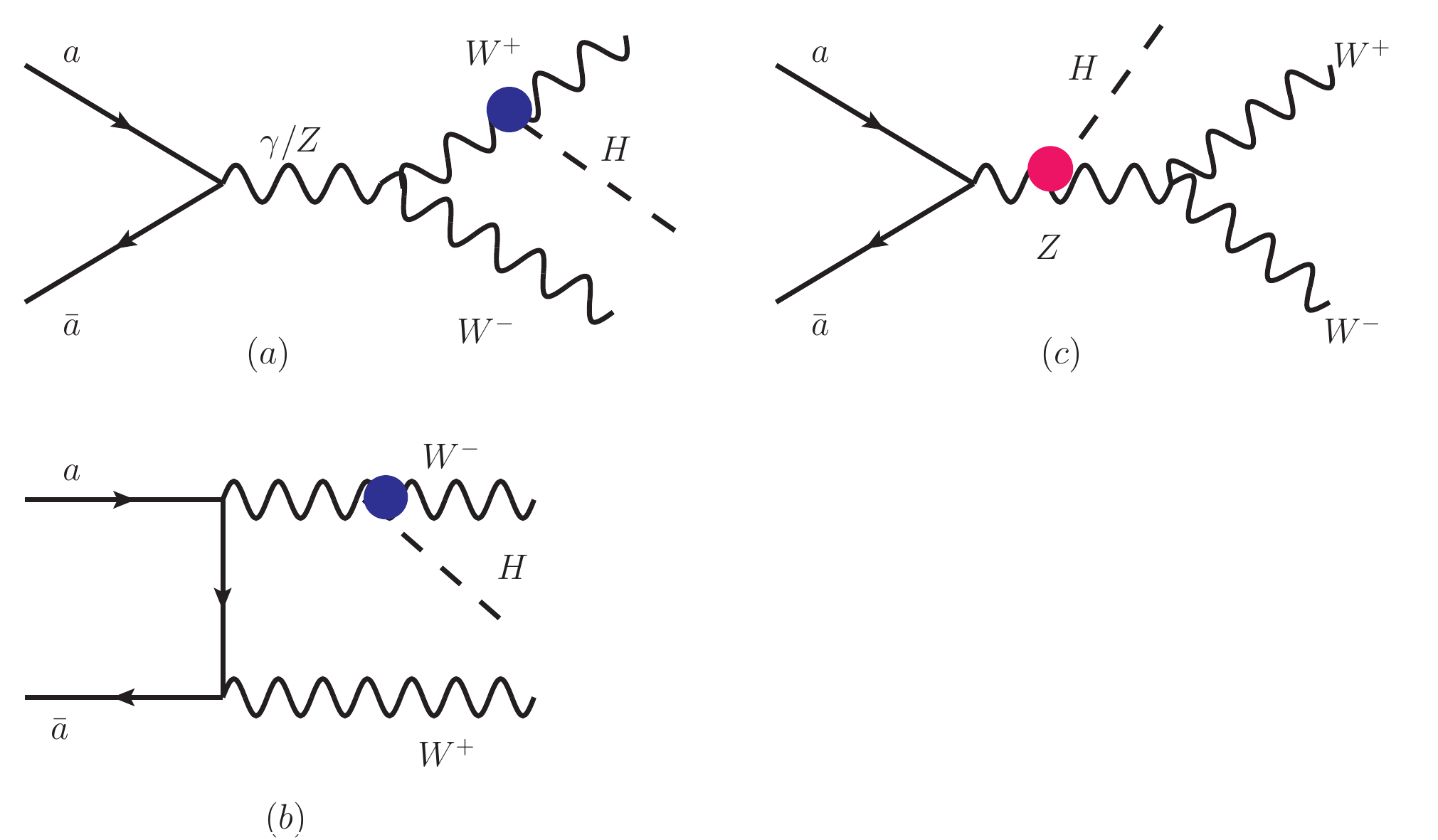}
\caption{ The representative Feynman diagrams of $a\bar{a}\to W^+W^-H$, where $a(\bar{a})$ denotes initial-state $q(\bar{q})$ or $e^-(e^+)$. The blue and red vertices represent the interactions of $HWW$ and $HZZ$, respectively. The squared amplitude of (a) and (b) with the SM $HWW$ coupling is denoted by $|\mathcal{M}_W|^2$ in Eq.~\eqref{eq:squared_amplitude}, and the squared amplitude of (c) with the SM $HZZ$ coupling is denoted by $|\mathcal{M}_Z|^2$, while the SM interference of (a) and (b) with (c) is $\mathcal{M}_{WZ}^2$. Here we have neglected the diagrams with small Yukawa couplings of the Higgs boson to light fermions.}
\label{fig:ePeM_wPwMh}
\end{figure}

While current measurements cannot determine the sign of $\lambda_{WZ}$, we propose that both its sign and magnitude can be directly measured through the tree-level interference in $W^+W^-H$ production. To show this, we first make a general discussion about the $W^+W^-H$ production at colliders.  At the parton level, the $a\bar{a}\to W^+W^-H$ process involves three types of Feynman diagrams shown in Fig.~\ref{fig:ePeM_wPwMh}, where $a(\bar{a})$ denotes $q(\bar{q})$ and $e^-(e^+)$ at hadron colliders and $e^+e^-$ colliders, respectively.
The squared amplitude of this process can be organized as:
\begin{align}
\label{eq:squared_amplitude}
|\mathcal{M}|^2 = |\mathcal{M}_W|^2 + |\mathcal{M}_Z|^2 + \mathcal{M}_{WZ}^2
~,
\end{align}
where ${\cal M}_W$ involves only the $HWW$ coupling, ${\cal M}_Z$ only the $HZZ$ coupling, and ${\cal M}_{WZ}$ the interference.  The total cross section of the production process can then be expressed as
\begin{align}
\label{eq:prod_cross_section}
\begin{split}
\sigma_{\text{prod}}
&=
\kappa_W^2 \left[ \sigma_W + \lambda_{WZ}^{-1} \sigma_{WZ} + \lambda_{WZ}^{-2} \sigma_{Z} \right],
\end{split}
\end{align}
where we have factored out the $\kappa$ factors so that $\sigma_{W,Z,WZ}$ represent the corresponding SM contributions. Here the interference term $\sigma_{WZ}$ is found to be negative. For the application to the case where $H$ is an exotic Higgs boson, one needs to change the mass of the Higgs boson in the evaluation of $\sigma_{W,Z,WZ}$.  Now that the total cross section is a function of $\kappa_W^2$ and $\lambda_{WZ}$, a measurement of it will determine $\lambda_{WZ}$ as a function of $|\kappa_W|$.  Combining with another experiment that measures $|\kappa_W|$ and/or $|\kappa_Z|$, it is then readily to determine $\lambda_{WZ}$.

In what follows, we will only make a brief comment on the $W^+W^-H$ production at the LHC because we find our proposed channel to be impractical in this case. It was claimed in Ref.~\cite{Gabrielli:2013era} that the $pp\to W^+W^-H$ process in the SM had the best sensitivity in the \llljj channel and  a  $3.0\sigma$ sensitivity can be reached at the HL-LHC with an integrated luminosity of $3\abi$. However, we find that the sensitivity is actually lower when other omitted factors are taken into account. First, Ref.~\cite{Gabrielli:2013era} considered the backgrounds of $WWZ$, $jjWZ$ with $Z\to \tau^+\tau^-$, where the $\tau$ lepton could further decay into $e$ or $\mu$, but ignored the backgrounds of $WWZ$, $jjWZ$ with $Z\to \ell^+\ell^-$ ($\ell=e,\mu$) under the assumption that they could be completely removed using the invariant mass cut of the same-flavor and opposite-sign (SFOS) lepton pairs.  Nevertheless, such a selection cut would also reduce the signal and other background events, with a net result of lowering the signal significance as we have checked.  Second, to reconstruct the $W$ boson, the cut $|m_{jj}-m_W|<5\gev$~\cite{Gabrielli:2013era} was imposed, which was the most important cut to suppress the backgrounds of $jjWWW$, $jjWZ$, $t\bar{t}W$, etc.  However, the efficiency of such a cut was somewhat overestimated~\cite{Aad:2015ona}.  Finally, jet matching was not considered in Ref.~\cite{Gabrielli:2013era}.  This might notably increase the backgrounds of $jjWWW$, $jjWZ$, etc. at hadron colliders.

Due to the above-mentioned low sensitivity of the $pp\to W^+W^-H$ process in the SM at the LHC, we reckon that it is not practical to use it to study the ratio $\lambda_{WZ}$. This leads us to the next section where we consider probing $\lambda_{WZ}$ through the $e^+ e^- \to W^+W^-H$ channel at future $e^+ e^-$ colliders.

\section{\tf{$W^+W^-H$}{WWH} production at future \tf{$e^+e^-$}{ee} colliders}
\label{sec:ILC}


\begin{figure}[!htb]
\centering
\includegraphics[width=0.5\textwidth]{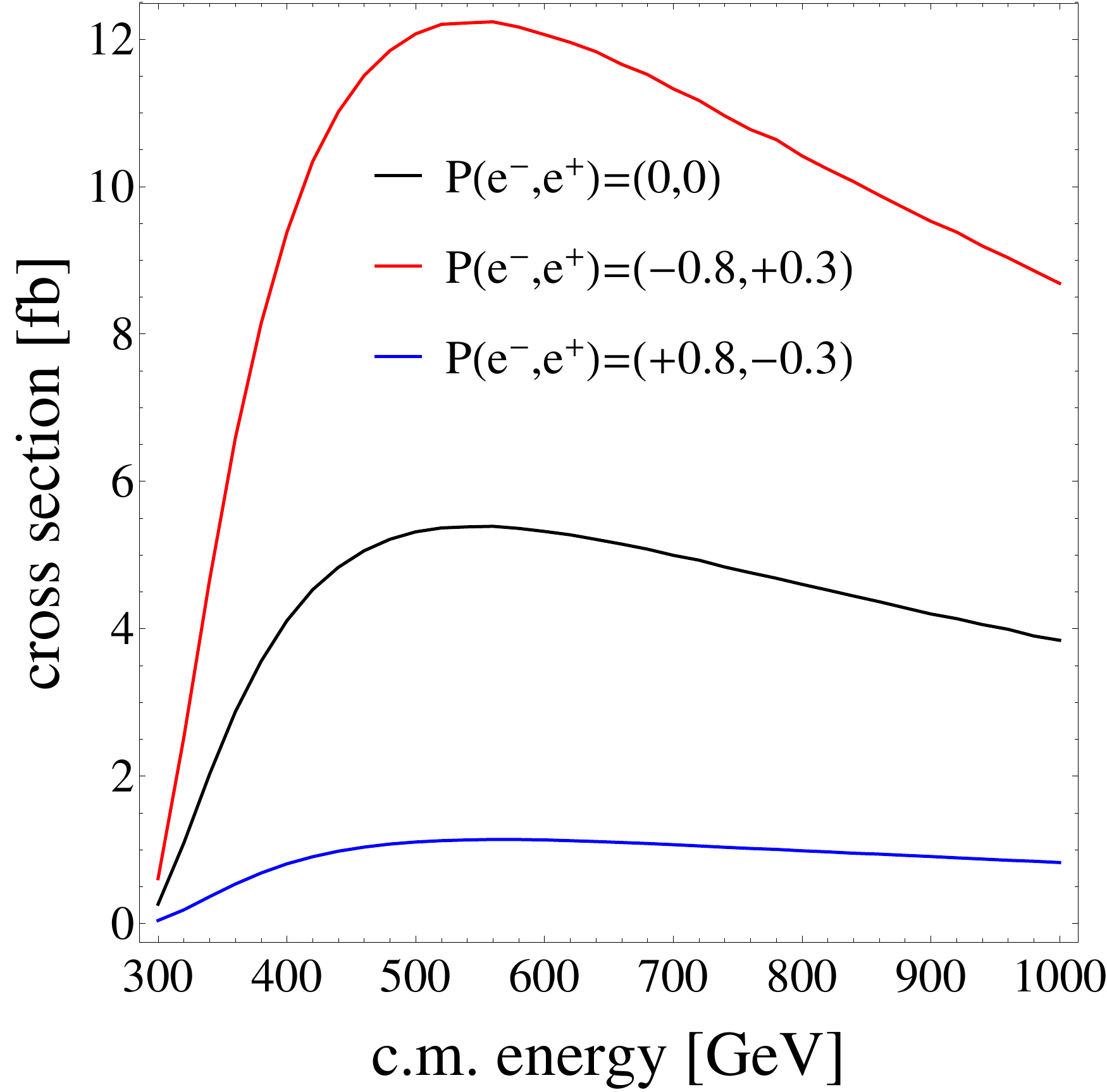}
\caption{Cross section of the \eewwh process as a function of c.m. energy for three choices of beam polarizations: $P(e^-,e^+)=(0,0)$, $(-0.8,+0.3)$ and $(+0.8,-0.3)$~\cite{Baer:2013cma}, where $m_H = 125$~GeV.}
\label{fig:scan_energy}
\end{figure}

The SM cross section of \eewwh as a function of the colliding energy at an $e^+e^-$ collider is shown in Fig.~\ref{fig:scan_energy}.  The blue, black and red curves are respectively for different polarization schemes: $P(e^-,e^+) = (+0.8,-0.3)$, $(0,0)$ and $(-0.8,+0.3)$.  It is observed that in all schemes the cross section reaches its maximum at $\sqrt{s}\sim 500 - 550\gev$~\cite{Kumar:2015eea}.  We note that the position of this maximum will change if $H$ represents an exotic Higgs boson with a mass different from 125~GeV.  The cross section increases by about a factor of 2 when the beam is polarized in the scheme of $P(e^-,e^+)=(-0.8,+0.3)$, as compared to the unpolarized case.  In view of the maximum position and to be specific about machine parameters, we consider a 500-GeV International Linear Collider (ILC) with the beam polarization of $P(e^-,e^+)=(-0.8,+0.3)$ and an integrated luminosity ${L}=4\abi$~\cite{Fujii:2015jha,Barklow:2015tja}~\footnote{In Ref.~\cite{Barger:1993wt}, as an example at the ILC, the integrated luminosities of $4\abi$ at $\sqrt{s}=500\gev$ is shared equally by the beam polarization choices of $P(e^-,e^+)=(-0.8,+0.3)$ and $P(e^-,e^+)=(+0.8,-0.3)$.  But it is also stressed that the polarization scheme should be modified depending on future experimental results~\cite{Barger:1993wt}. For other studies with $P(e^-,e^+)=(-0.8,+0.3)$ and the maximal integrated luminosity of $4\abi$ at the 500-GeV ILC, see for example Ref.~\cite{Chiu:2017yrx}. }.  The tree-level interference effect in the \eewwh process can also be studied at other future $e^+e^-$ colliders, provided the colliding energy is sufficiently high~\cite{Gomez-Ceballos:2013zzn,Aicheler:2012bya}.

To study the process \eewwh with general $\kappa_W$ and $\kappa_Z$, we first consider the following benchmark scenarios:
\begin{align}
\begin{split}
&\text{BP1:}\quad \kappa_W=1,\ \kappa_Z=1,
\\
&\text{BP2:}\quad \kappa_W=1,\ \kappa_Z=-1,
\\
&\text{BP3:}\quad \kappa_W=1,\ \kappa_Z=0.
\end{split}
\end{align}
The corresponding total cross sections are
\begin{align}
\sigma_{\text{BP1}}=12\fb,\quad
\sigma_{\text{BP2}}=17.11\fb,\quad
\sigma_{\text{BP3}}=13.54\fb.
\end{align}
Thus we obtain each term in Eq.~\eqref{eq:prod_cross_section} in this case as follows:
\begin{align}
\sigma_{W}=13.54\fb,\quad \sigma_{Z}=1.015\fb,\quad \sigma_{WZ}=-2.555\fb.
\end{align}
The diagrams involving the $HWW$ coupling dominate the one involving the $HZZ$ coupling by about one order of magnitude, and the interference between the two types of diagrams is destructive.

Two decay channels $H\to b\bar{b}$ and $H\to WW^*$ are considered with the corresponding final states being \jjlbb and \llljj, respectively.  Let us now define $\epsilon_{\text{BP1}}$, $\epsilon_{\text{BP2}}$ and $\epsilon_{\text{BP3}}$ to be the efficiencies of the signals under a fixed set of cuts for the BP1, BP2, and BP3 scenarios, respectively.  The signal cross section with arbitrary $HWW$ and $HZZ$ couplings after the selection cuts can now be written as
\begin{align}
\label{eq:signal_xsec}
\sigma_{S}
=
\kappa_W^2 \left( 
\sigma_{W}\epsilon_W + \lambda_{WZ}^{-1} \sigma_{WZ}\epsilon_{WZ} + \lambda_{WZ}^{-2} \sigma_{Z} \epsilon_Z
\right) \mathcal{B},
\end{align}
where $\mathcal{B}=\mathcal{B}(\kappa_W,\kappa_Z)$ denotes the combined branching ratios of the $W$ and $H$ decays in the final state and depends on the coefficients $\kappa_W$ and $\kappa_Z$, and $\epsilon_W$, $\epsilon_Z$ and $\epsilon_{WZ}$ denote the cut efficiencies of the associated parts.  One can then solve to obtain
\begin{align}
\epsilon_{W}&=\dfrac{\sigma_{\text{BP3}}\epsilon_{\text{BP3}}}{\sigma_W},\nn\\
\epsilon_{WZ}&=\dfrac{\sigma_{\text{BP1}}\epsilon_{\text{BP1}}-
\sigma_{\text{BP2}}\epsilon_{\text{BP2}}}{2\sigma_{WZ}},\nn\\
\epsilon_{Z}&=\dfrac{\sigma_{\text{BP1}}\epsilon_{\text{BP1}}-
\sigma_{\text{BP3}}\epsilon_{\text{BP3}}}{\sigma_Z}
-\dfrac{\sigma_{\text{BP1}}\epsilon_{\text{BP1}}-
\sigma_{\text{BP2}}\epsilon_{\text{BP2}}}{2\sigma_{Z}}.
\end{align}
Therefore, through a simulation for the three benchmark scenarios, we are able to extract $\epsilon_W$, $\epsilon_Z$ and $\epsilon_{WZ}$ for general analyses.

In this work, we perform a parton-level simulation with the signal and background events generated using \texttt{MG5\_aMC@NLO v2.4.3}~\cite{Alwall:2014hca}. In order to mimic detector resolution effects, particle four-momenta are smeared with a Gaussian distribution.  The jet energy resolution and lepton momentum resolution are approximately described by~\cite{Abe:2010aa}~
\footnote{It should be noted that although the energy of an electron can be measured in the electromagnetic calorimeter, the measurement is not statistically independent of the tracking determination of its momentum so that these two measurements cannot be combined~\cite{Gunion:1998jc}. In practice, we smear the electron momentum with track system performance in Eq.~\eqref{eq:track} as in Refs.~\cite{Liu:2016zki,Liu:2017zdh}.
}
\begin{align}
\dfrac{\Delta E}{E}
&=
\dfrac{0.3}{\sqrt{E/\gev}},\\
\label{eq:track}
\Delta \left( \dfrac{1}{p_T} \right) 
&= 
2\times 10^{-5}\oplus\dfrac{10^{-3}}{p_T\sin\theta},
\end{align}
respectively.

\subsection{The \tf{$H\to b\bar{b}$}{Hbb} case}

We first consider the signal process \eewwh in the $jj\ell^{\pm}bb$ channel, where $j$ denotes a light-flavored quark jet, $H\to b\bar{b}$, and one of the $W$ bosons decays hadronically while the other decays leptonically.  The main backgrounds include $t\bar{t}$ and $WWZ$ with $Z\to b\bar{b}, jj$. The following basic cuts are imposed at the generator level:
\begin{align}
\label{eq:basic_cuts}
p_{T}^{j}>20\gev,\quad p_{T}^{\ell}>10\gev,\quad |\eta_{\ell,j}|<2.5,\quad \Delta R_{mn}>0.4,\quad m,n=j,\ell,
\end{align}
where the angular distance in the $\eta-\phi$ plane $\Delta R_{ij}\equiv\sqrt{(\eta_i-\eta_j)^2+(\phi_i-\phi_j)^2}$, and $\eta_i$ and $\phi_i$ are the pseudorapidity and azimuthal
angle of particle $i$, respectively.

\begin{table}[t!hb]
\tabcolsep=12pt
\caption{Cut flow of signal and background cross sections in the $jj\ell^{\pm}bb$ channel.}
\begin{tabular}{c|cccc}
\hline\hline
cross section (fb)& basic cuts, $b$-tagged& $m_{jj}$ & $m_{bb}$ & $p_{T}^{\ell}$ \\
\hline
BP1         & 0.721	&	0.703	&	0.702	&	0.581
\\ 
BP2         & 1.06	&	1.03		&	1.03		&	0.864
\\ 
BP3         & 0.822	&	0.802	&	0.795	&	0.664
\\ \hline
$t\bar{t}$  & 87.5	&	85.6		&	11.0		&	7.92
\\ 
$WWZ$       & 1.09	&	1.07		&	0.0305	&	0.023
\\ 
\hline\hline
\end{tabular}
\label{tbl:ILC_hbb}
\end{table}

We assume that the $b$-tagging efficiency, the rates of misidentifying the $c$-jet and the light-quark jet as a $b$-jet are 0.8, 0.08 and 0.01, respectively~\cite{Suehara:2015ura}. 
To reconstruct the $W$ and $H$ bosons, we require that the invariant mass of the light jet pair and the $b$-jet pair satisfies~\cite{Durig:2014lfa}~
\footnote{The asymmetric cut on the invariant mass of $b\bar{b}$ is required to take into account the effects of photons from initial-state radiation (ISR)~\cite{Durig:2014lfa} or unmeasured neutrinos from semi-leptonic $b$-decays~\cite{Aad:2014yja}.}
\begin{align}
|m_{jj}-m_{W}|<15\gev,\quad m_{H}-20\gev < m_{bb} < m_{H}+10\gev,
\end{align}
where the Higgs boson mass $m_{H}=125\gev$ here.  To suppress the $t\bar{t}$ background, we further require $p_{T}^{\ell}>30\gev$. The cut flow of the signals in scenarios (BP1, BP2, BP3) and backgrounds are summarized in Table~\ref{tbl:ILC_hbb}.  It is observed that the $m_{bb}$ cut can effectively reduce both backgrounds.

Here the combined decay branching ratio
\begin{align}
\mathcal{B}=\mathcal{B}_0\dfrac{1}{\Gamma_{H}/\Gamma_{H}^{0}},
\end{align}
where $\mathcal{B}_0=16.4\%$ is the corresponding value for the SM Higgs boson~\cite{Patrignani:2016xqp} and the total width of $H$ is~\footnote{The modification of the partial width $H\to \gamma\gamma$ from $\kappa_W$ is much smaller than that of the partial width $H\to W^+W^-$ and thus neglected in Eq.~\eqref{eq:totwidth_1}. }
\begin{align}
\label{eq:totwidth_1}
\Gamma_{H}=\Gamma_{H}^{0}\big[1+(\kappa_W^2-1)\mathcal{B}_{HWW}^{0}
+(\kappa_Z^2-1)\mathcal{B}_{HZZ}^{0}\big],
\end{align}
with the SM decay branching ratios and total width of the Higgs boson given by $\mathcal{B}_{HWW}^{0}=0.214$, $\mathcal{B}_{HZZ}^{0}=2.62\%$ and $\Gamma_{H}^{0}=4.07$~MeV~\cite{Patrignani:2016xqp}.  We have assumed that the couplings of $H$ to the SM fermions have the same values as in the SM.

The cut efficiencies are the ratios of the cross sections in the last column in Table~\ref{tbl:ILC_hbb} to the corresponding cross sections with no cut. For the three benchmark scenarios, they are found to be
\begin{align}
\epsilon_{\text{BP1}}=0.296\quad
\epsilon_{\text{BP2}}=0.308\quad
\epsilon_{\text{BP3}}=0.300.
\end{align}
Therefore, we obtain the cut efficiency for each of the contributions in Eq.~\eqref{eq:signal_xsec} as 
\begin{align}
\epsilon_W=0.300, \quad \epsilon_Z=0.346,\quad \epsilon_{WZ}=0.337.
\end{align}
We then evaluate the discovery significance using~\cite{Cowan:2010js}
\begin{align}
S_D&=\sqrt{2\big[(n_s+n_b)\text{ln}\dfrac{n_s+n_b}{n_b}-n_s\big]},
\end{align}
where the number of signal and background events after selection cuts are
\begin{align}
n_s=\sigma_S {L} ~,~  n_b=(\sigma_{t\bar{t}}+\sigma_{WWZ}) {L} ~,
\end{align}
with ${L}$ being the integrated luminosity, and $\sigma_{t\bar{t}}$, $\sigma_{WWZ}$ the corresponding cross sections of the backgrounds of $t\bar{t}$, $WWZ$ with $Z\to b\bar{b},jj$ after the selection cuts.

\begin{figure}[t!hb]
\centering
\includegraphics[width=0.45\textwidth]{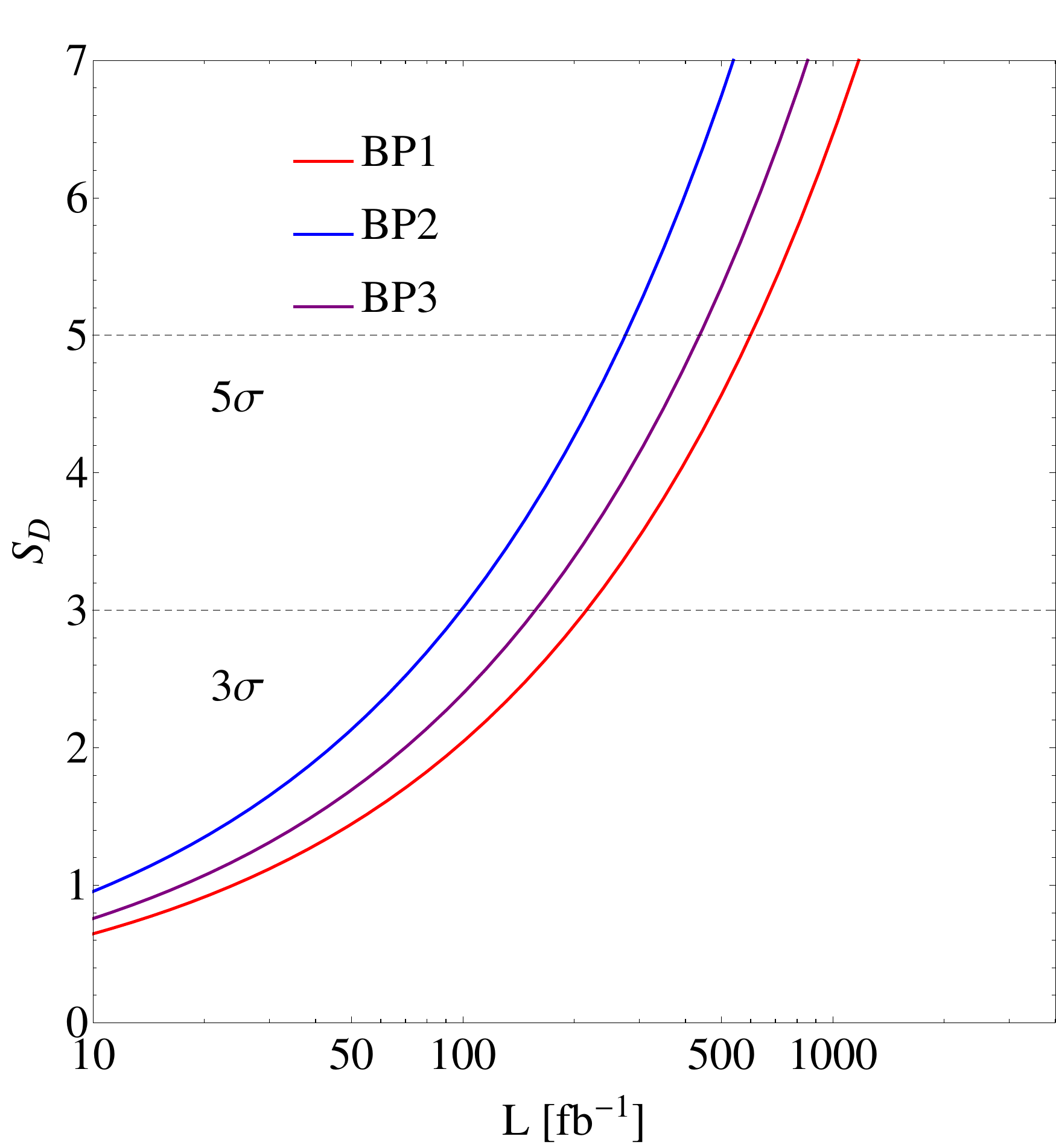}
\hspace{5mm}
\includegraphics[width=0.45\textwidth]{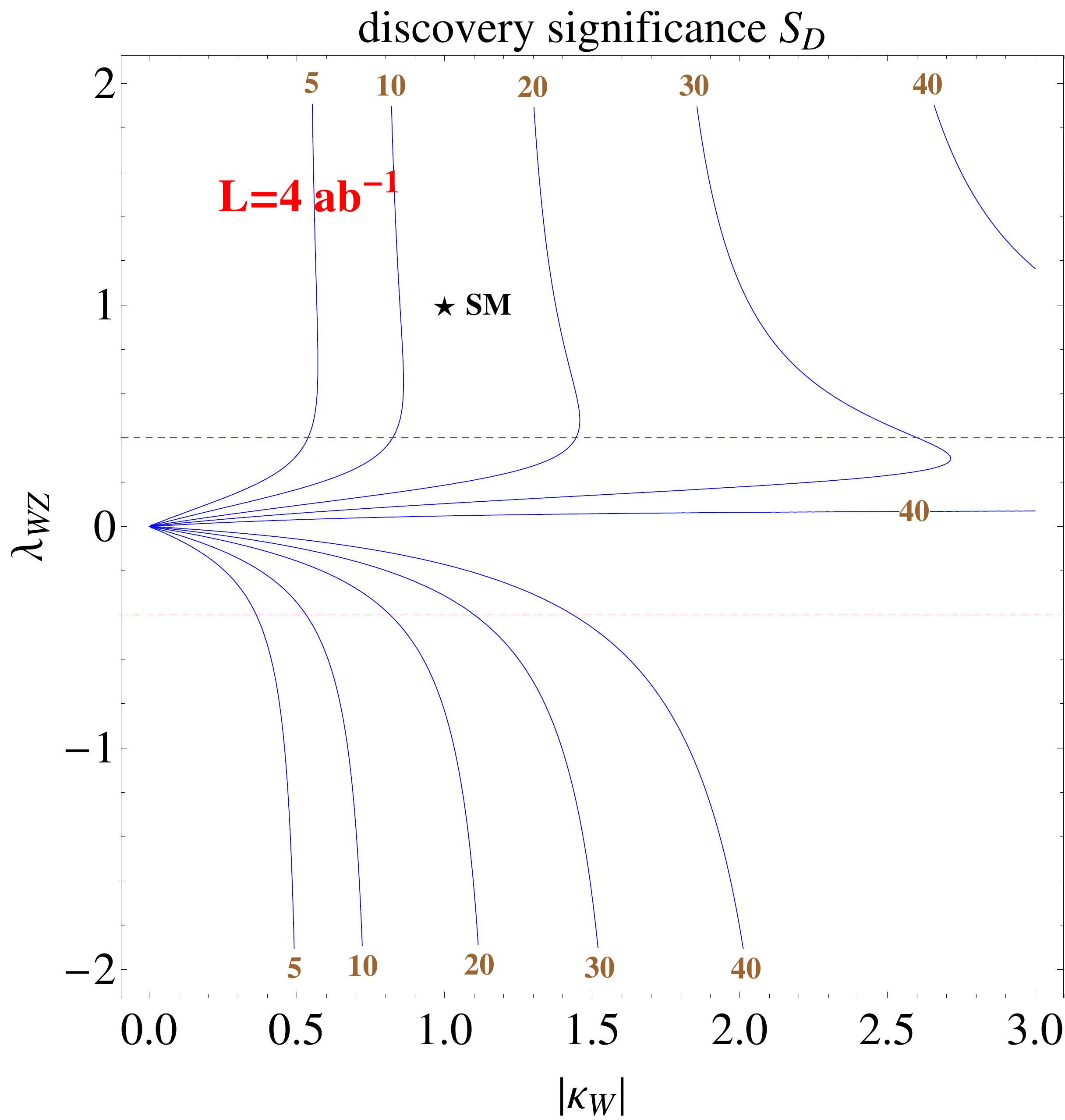}
\caption{Discovery potential of the $e^+e^-\to W^+W^-H$ process in the $jj\ell^{\pm}bb$ channel at the 500-GeV ILC with the beam polarization $P(e^-,e^+)=(-0.8,+0.3)$.  Left: the signal significance of the three benchmark scenarios defined in the main text as a function of the integrated luminosity.  Right: contours of signal significance in the $|\kappa_W|$-$\lambda_{WZ}$ plane, assuming an integrated luminosity of $4\abi$.  The red dashed lines correspond to $\lambda_{WZ}=\pm 0.4$.  The star ($\star$) marks the SM scenario (BP1).}
\label{fig:discovery_hbb}
\end{figure}

\begin{figure}[!htb]
\centering
\includegraphics[width=0.45\textwidth]{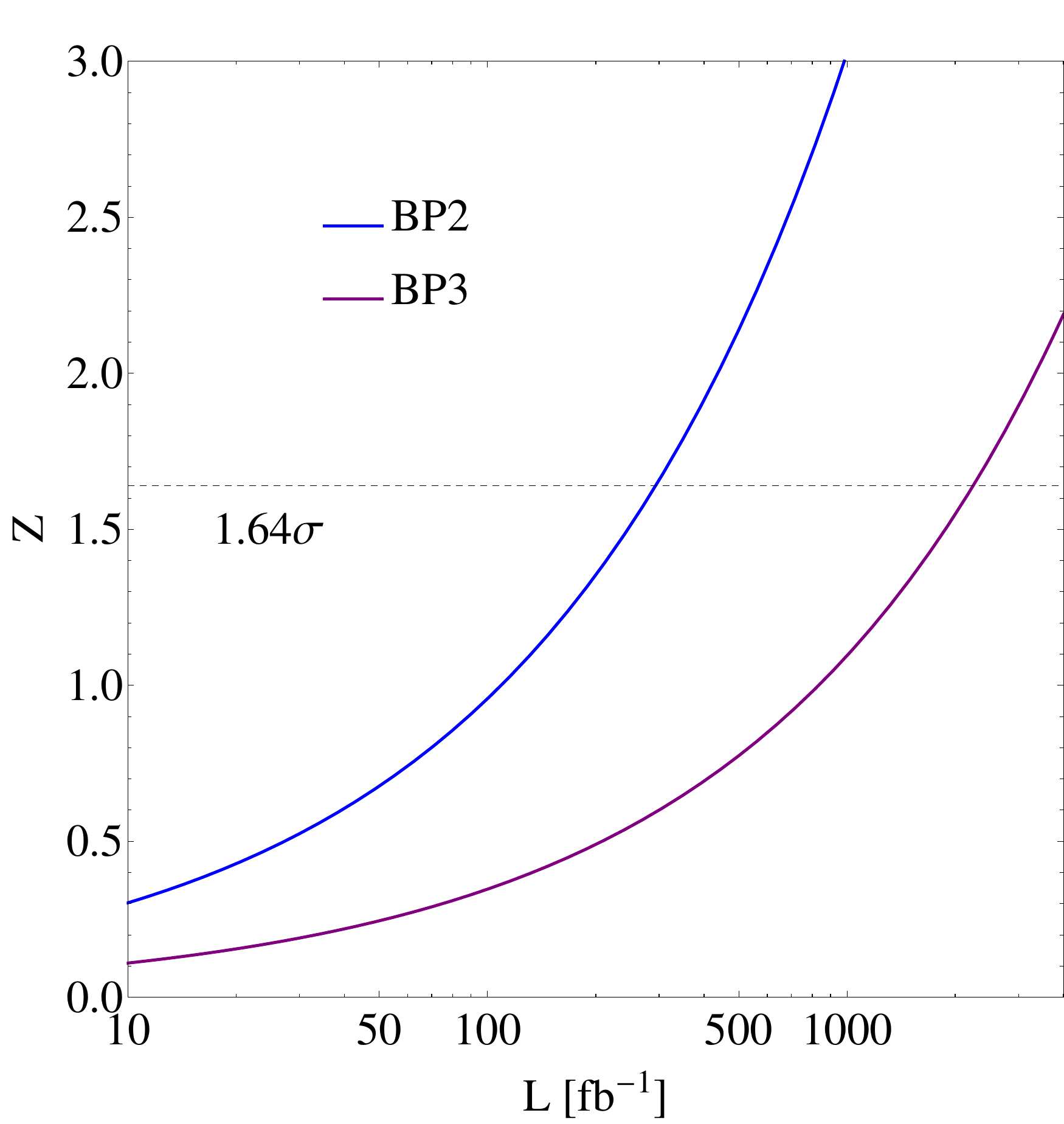}
\hspace{5mm}
\includegraphics[width=0.45\textwidth]{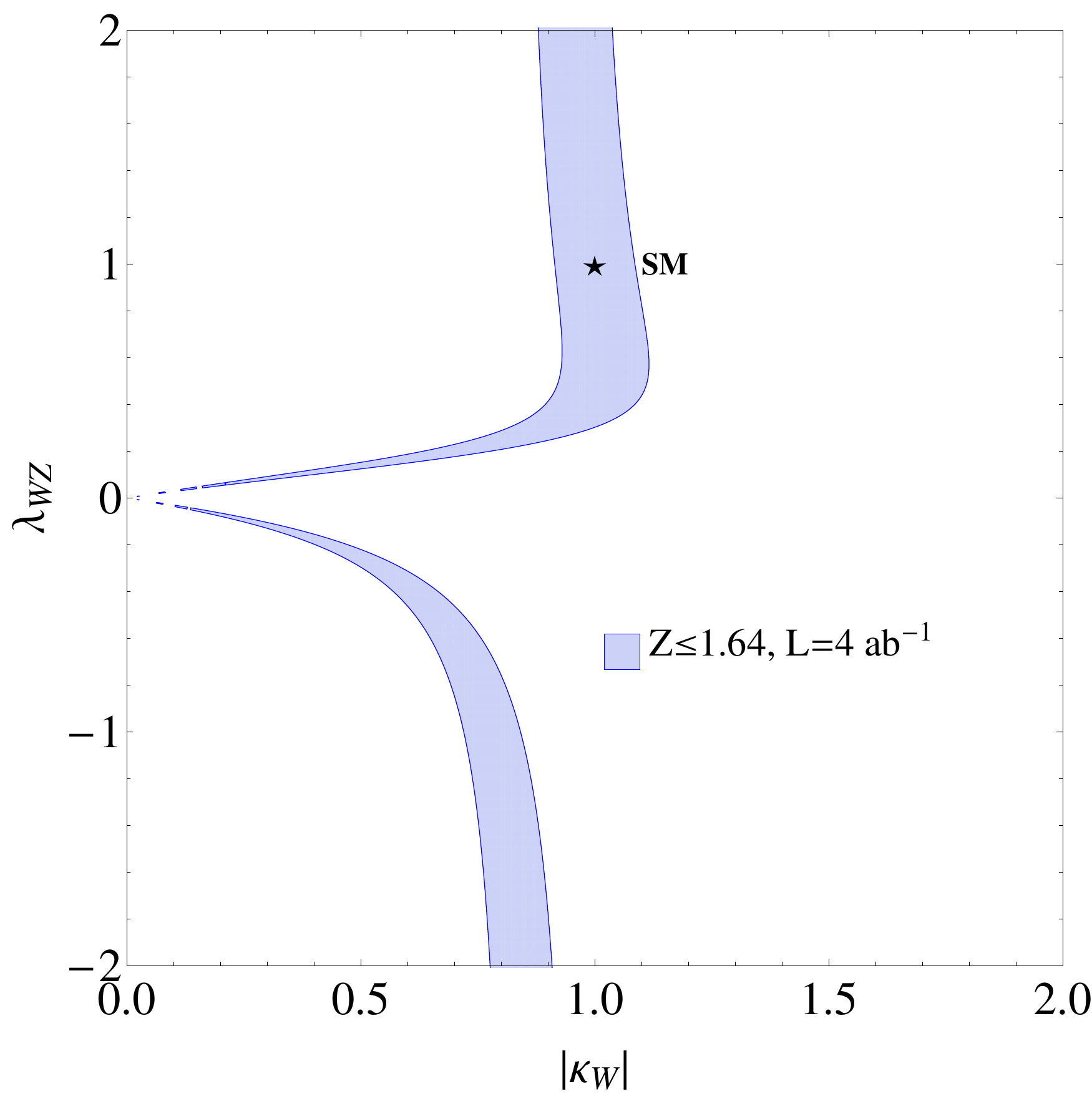}
\caption{Power to discriminate $\lambda_{WZ}$ in the $jj\ell^{\pm}bb$ channel at the 500-GeV ILC with beam polarization $P(e^-,e^+)=(-0.8,+0.3)$.  Left: discriminating the (BP2, BP3) scenarios from the SM scenario (BP1).  Right: region in the $|\kappa_W|$-$\lambda_{WZ}$ plane that satisfies $Z \le 1.64$, assuming the integrated luminosity of $4\abi$. }
\label{fig:exclusion_hbb}
\end{figure}

Fig.~\ref{fig:discovery_hbb} shows the signal significance of the $e^+e^-\to W^+W^-H$ process in the $jj\ell^{\pm}bb$ channel at the 500-GeV ILC with the beam polarization scheme of $P(e^-,e^+)=(-0.8,+0.3)$.  The left plot considers the three scenarios (BP1, BP2, BP3).  It is seen that a $5\sigma$ discovery can be achieved with an integrated luminosity of about $600\fbi$, $300\fbi$ and $450\fbi$ for (BP1, BP2, BP3), respectively.  The SM scenario (BP1) requires a larger luminosity because it has the smallest cross section among them.  We also note that for a fixed $|\kappa_W|$,  the production cross section takes its minimum when $\lambda_{WZ} \simeq 0.82$. The right plot shows the contours of signal significance in the $|\kappa_W|$-$\lambda_{WZ}$ plane, assuming an integrated luminosity of $4\abi$.  It is found that the \eewwh process can be discovered in the $jj\ell^{\pm} bb$ channel if $|\kappa_W|\gtrsim 0.6$, irrespective of the value of $\lambda_{WZ}$.  Besides, the \eewwh process is more sensitive to scenarios with $\lambda_{WZ} \alt 0.41$ because the destructive interference contribution $\sigma_{WZ}$ in Eq.~\eqref{eq:prod_cross_section} becomes less important than the $\sigma_Z$ contribution.


To distinguish the SM hypothesis $\lambda_{WZ}=1$ from a non-SM hypothesis $\lambda_{WZ}=\lambda_0(\neq 1)$, we define the test ratio as~\cite{Cowan:2010js}~
\footnote{See Refs.~\cite{Dutta:2017sod,Harnik:2013aja,Askew:2015mda} for similar test ratios in other studies.}, 
\begin{align}
\label{eq:likelihood}
Q=-2 \ln \dfrac{L(\lambda_{0})}{L(1)},
\end{align}
where the likelihood function $L(\lambda_{WZ})$ is defined as 
\begin{align}
L(\lambda_{WZ})
=
P(\text{data} | n_b+n_s^{\lambda_{WZ}}) ~,
\end{align}
with the Poisson distribution function given by
\begin{align}
P(k|\lambda)=\lambda^k e^{-\lambda}/k! 
~.
\end{align}
We can calculate the $p$-value of the non-SM hypothesis using the likelihood ratio in Eq.~\eqref{eq:likelihood} by assuming that the actual observation is taken to be the median of the $Q$ distribution under the SM hypothesis~\cite{Cowan:2010js}, {\it i.e.},
\begin{align}
\text{data}=n_b+n_s^{\lambda_{WZ}=1}.
\end{align}
The non-SM hypothesis $\lambda_{WZ}=\lambda_0$ is then rejected at the $Z$-sigma level with $Z=\Phi(1-p)^{-1}$, where $\Phi$ is the cumulative distribution of the standard Gaussian~\cite{Cowan:2010js}. Explicitly,
\begin{align}
\label{eq:z_value}
Z=\sqrt{2\left[ \left( n_s^{\lambda_{WZ}=1}+n_b\right) \text{ln}\dfrac{n_s^{\lambda_{WZ}=1}+n_b}{n_s^{\lambda_{WZ}=\lambda_0}+n_b}+\left( n_s^{\lambda_{WZ}=\lambda_0}-n_s^{\lambda_{WZ}=1}\right) \right] }.
\end{align}

\begin{figure}[!htb]
\centering
\includegraphics[width=0.5\textwidth]{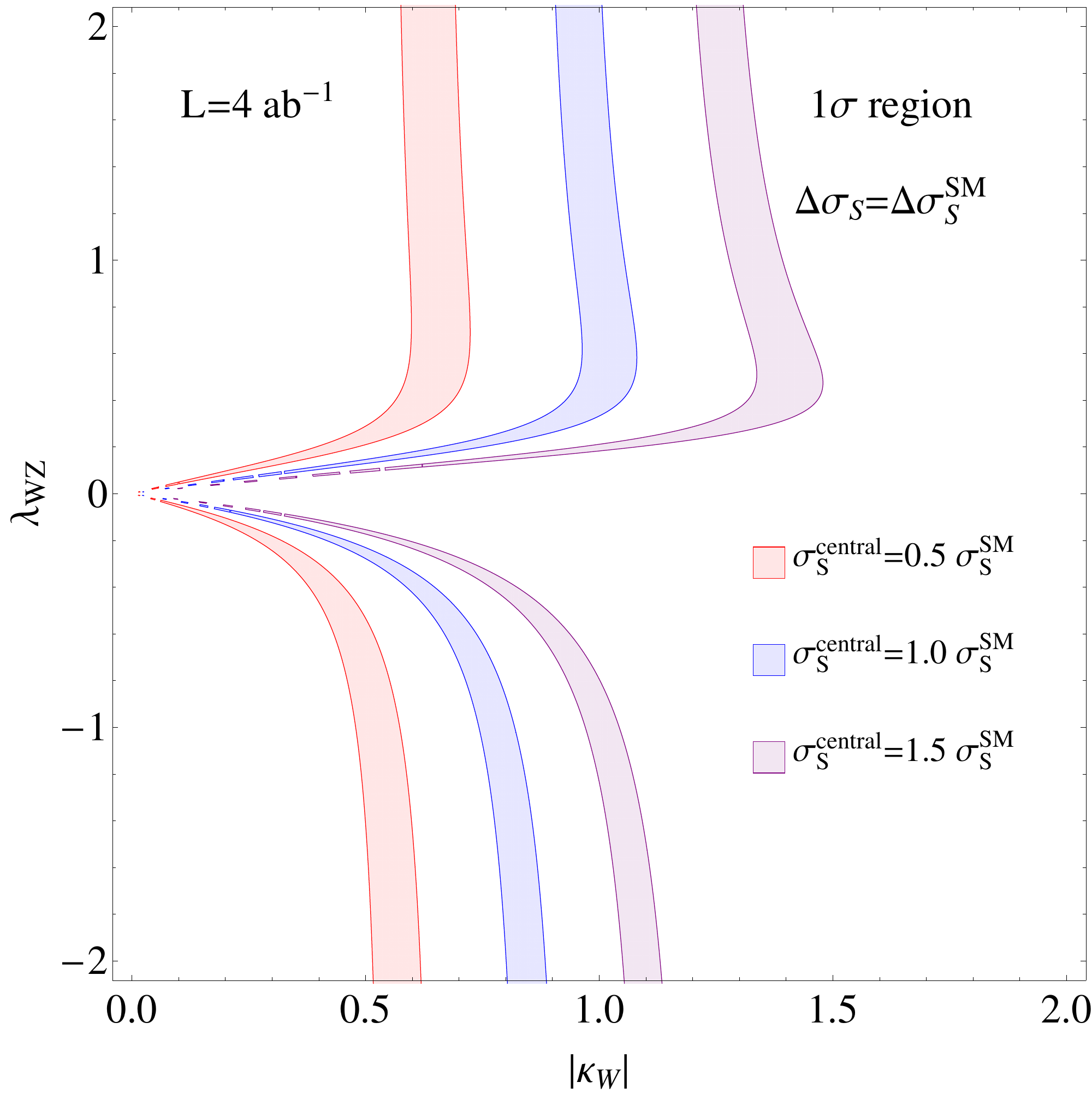}
\caption{$1\sigma$ regions in the $|\kappa_W|$-$\lambda_{WZ}$ plane when the measured signal cross section has the central value $\sigma_S^{\text{central}}= (0.5, 1, 1.5) \times \sigma_S^{\text{SM}}$ and an error bar assumed to be the same as the SM expectation, assuming the 500~GeV ILC with the beam polarization $P(e^-,e^+)=(-0.8,+0.3)$ and the integrated luminosity of $4\abi$.}
\label{fig:precision_hbb}
\end{figure}

In Fig.~\ref{fig:exclusion_hbb}, we show the power to discriminate $\lambda_{WZ}$ from its SM value.  The left plot shows that the (BP2, BP3) scenarios can be rejected from the SM (BP1) at 95\% C.L. ({\it i.e.}, $Z \ge 1.64$) when the integrated luminosity goes above $290\fbi$ and $2.2\abi$, respectively. 
The right plot, on the other hand, shows the region in the $|\kappa_W|$-$\lambda_{WZ}$ plane that satisfies $Z \le 1.64$ with an integrated luminosity of $4\abi$.  We thus find that if $|\kappa_W|\gtrsim 0.9$ as determined by other measurements, a good portion of negative $\lambda_{WZ}$ region can be completely excluded at 95\% C.L. through the \eewwh process in the $jj\ell^{\pm}bb$ channel.

As the central value of observed \eewwh cross section may be different from the SM value, we consider three different cases for measured central value:
\begin{align}
\label{eq:central_values}
\sigma_S^{\text{central}} 
= 
(0.5, 1, 1.5) \times \sigma_S^{\text{SM}}
~,
\end{align}
where $\sigma_S^{\text{SM}}$ denotes the central value of SM expectation, and keep their statistical errors the same as the SM one, which is~
\footnote{The definition of $1\sigma$ statistical error using the log likelihood ratio can be found in Ref.~\cite{Han:2015ofa}.}
\begin{align}
\label{eq:stat_err}
\Delta\sigma_S^{\text{SM}}
=
\dfrac{\sqrt{n_s^{\lambda_{WZ}=1}+n_b}}{n_s^{\lambda_{WZ}=1}} \sigma_S^{\text{SM}}
\simeq 0.082 \sigma_S^{\text{SM}}
\end{align}
in the $jj\ell^{\pm}bb$ channel with an integrated luminosity of $4\abi$.  We note that for $|\kappa_W| = 1$, the predicted signal cross section cannot be lower than about $0.97\sigma_S^{\text{SM}}$.
Fig.~\ref{fig:precision_hbb} shows allowed parameter regions using the above assumed experimental results.  Depending on the measurement outcome and the information of $|\kappa_W|$ and/or $|\kappa_Z|$ from other experiments, one can readily fix the value, particularly the sign, of $\lambda_{WZ}$.  For example, if $|\kappa_W|$ is determined to be around 1 and $\sigma_S^{\text{central}}=\sigma_S^{\text{SM}}$, then a positive $\lambda_{WZ}$ is favored. 

Since the signal cross section depends on $\kappa_W \, (\kappa_Z)$ and $\lambda_{WZ}$, we can obtain the relative error on $\lambda_{WZ}$ if that on $\kappa_W \, (\kappa_Z)$ is known.  As discussed in Section~\ref{sec:constraint}, the projected $1\sigma$ relative errors on $\kappa_W$ and $\kappa_Z$ at the HL-LHC are $\delta \kappa_W/\kappa_W < 0.05$ and $\delta \kappa_Z/\kappa_Z < 0.04$.  Moreover,
\begin{eqnarray}
\label{eq:error_propagation}
{\delta \lambda_{WZ}/ \lambda_{WZ}} = \sqrt{{(\delta \sigma_S)^2 + (\partial \sigma_S/ \partial\kappa_Z )^2 (\delta\kappa_Z)^2
\over (\partial \sigma_S/\partial\lambda_{WZ})^2\lambda_{WZ}^2}}\;,
\end{eqnarray}
where $\delta\sigma_S$ is equal to $\Delta\sigma_S^{\text{SM}}$, which is defined in Eq.~\eqref{eq:stat_err}. Therefore, we obtain $\delta\lambda_{WZ}/\lambda_{WZ} < 0.063$ for the SM scenario (BP1) with the combination of the \eewwh process and the $|\kappa_Z|$ measurement at the HL-LHC, which is even better than that with the combination of $|\kappa_W|$ and $|\kappa_Z|$ measurements at the HL-LHC as shown in Eq.~\eqref{eq:precision_lam}.  Alternatively, we would have $\delta\lambda_{WZ}/\lambda_{WZ} < 1.15$ if the $|\kappa_W|$ measurement is used instead. This is because $\partial \sigma_S/\partial\lambda_{WZ}$ with $\sigma_S$ in terms of $|\kappa_Z|$ and $\lambda_{WZ}$ is larger than that in terms of $|\kappa_W|$ and $\lambda_{WZ}$.

\subsection{The \tf{$H\to W^+W^-$}{HWW} case}

In this subsection, we consider the signal process \eewwh in the $\ell^{\pm}\ell^{\pm}\ell^{\mp} jj$ channel.  For the SM Higgs boson, the sensitivity of this channel is not competitive with that of the $jj\ell^{\pm}bb$ channel.  However, for an exotic Higgs boson with mass close to the SM Higgs boson and not coupling to the SM fermions at tree level, such as the fiveplet $H_5^0$ in the Georgi-Machacek (GM) model~\cite{Georgi:1985nv,Chanowitz:1985ug,Chiang:2012cn,Chiang:2015kka,Chiang:2015rva}, this channel provides a unique signature.  For a heavy exotic Higgs boson with the same couplings to the SM particles as the SM Higgs boson, this channel is also expected to become more dominant until the $t \bar t$ channel becomes open.  In the following, we study two schemes: (A) the couplings of $H$ to the SM fermions are the same as the SM Higgs boson, and (B) $H$ does not couple to the SM fermions at tree level.

Main backgrounds here include $4W$, $WWZ$ with $Z\to \ell^+\ell^-$, $\tau^+\tau^-$, where the $\tau$ lepton can further decay into $e$ or $\mu$.  We impose the same basic cuts as those in Eq.~\eqref{eq:basic_cuts}.  In order to reconstruct the $W$ boson and suppress the background with $Z$ boson decays, we require that~\cite{Durig:2014lfa,Aad:2015ona}
\begin{align}
|m_{jj}-m_W|<15\gev,\quad
|m_{\ell^+\ell^-}^{\text{SFOS}}-m_Z|>25\gev.
\end{align} 
The cut flow of the signals in scenarios (BP1, BP2, BP3) and backgrounds are summarized in Table~\ref{tbl:ILC_hww}.  The $m_{\ell^+\ell^-}^{\text{SFOS}}$ cut is seen to be effective in reducing the dominant background of $WWZ(\ell^+\ell^-)$.

\begin{table}[!htb]
\tabcolsep=12pt
\caption{Cut flow of the signal and background cross sections in the $\ell^{\pm}\ell^{\pm}\ell^{\mp} jj$ channel.}
\begin{tabular}{l|ccc}
\hline\hline
cross section (ab) & basic cuts & $m_{jj}$ & $m_{\ell^+\ell^-}^{\text{SFOS}}$ \\
\hline
BP1 & 25.5 & 22.4 & 16.9
\\ 
BP2 & 36.6 & 32.4 & 24.6
\\ 
BP3 & 28.4 & 25.0 & 18.9
\\ 
\hline
$4W$ & 4.92 & 4.57 & 3.04
\\ 
$WWZ(\ell^+\ell^-)$ & 919 & 896 & 7.59
\\ 
$WWZ(\tau^+\tau^-)$ & 9.53 & 9.35 & 6.85
\\ 
\hline\hline
\end{tabular}
\label{tbl:ILC_hww}
\end{table}

Suppose the total width of $H$ is denoted by $\Gamma_H$ and, again, the SM Higgs boson width is denoted by $\Gamma_{H}^0$.  The combined decay branching ratio in either scheme is given by:
\begin{align}
\label{eq:br_hww}
\mathcal{B}=\mathcal{B}_0\dfrac{\kappa_W^2}{\Gamma_{H}/\Gamma_{H}^{0}},
\end{align}
similar to that in the $H\to b\bar{b}$ decay channel. The SM value of the combined branching ratio is $\mathcal{B}_0=0.578\%$~\cite{Patrignani:2016xqp}.  The total width of $H$~
\footnote{We only consider the tree-level decays of $H$ for scheme B. }
\begin{equation}
\Gamma_{H}=
\begin{cases}
\Gamma_{H}^{0}\left[1+\left(\kappa_W^2-1\right)\mathcal{B}_{HWW}^{0}
+\left(\kappa_Z^2-1\right)\mathcal{B}_{HZZ}^{0}\right], &\mbox{for scheme A,}\\
\Gamma_{H}^{0}\left[\kappa_W^2\mathcal{B}_{HWW}^{0}
+\kappa_Z^2\mathcal{B}_{HZZ}^{0}\right], &\mbox{for scheme B.}
\end{cases}
\end{equation}

Through simulations, the cut efficiencies of the benchmark scenarios are found to be
\begin{align}
\epsilon_{\text{BP1}}=0.243,\quad
\epsilon_{\text{BP2}}=0.250,\quad
\epsilon_{\text{BP3}}=0.242.
\end{align}
Hence, we obtain the cut efficiencies of each contribution in Eq.~\eqref{eq:signal_xsec} as
\begin{align}
\epsilon_{W}=0.242,\quad
\epsilon_{Z}=0.309,\quad
\epsilon_{WZ}=0.264.
\end{align}

\subsubsection{Scheme A}

\begin{figure}[!htb]
\centering
\includegraphics[width=0.45\textwidth]{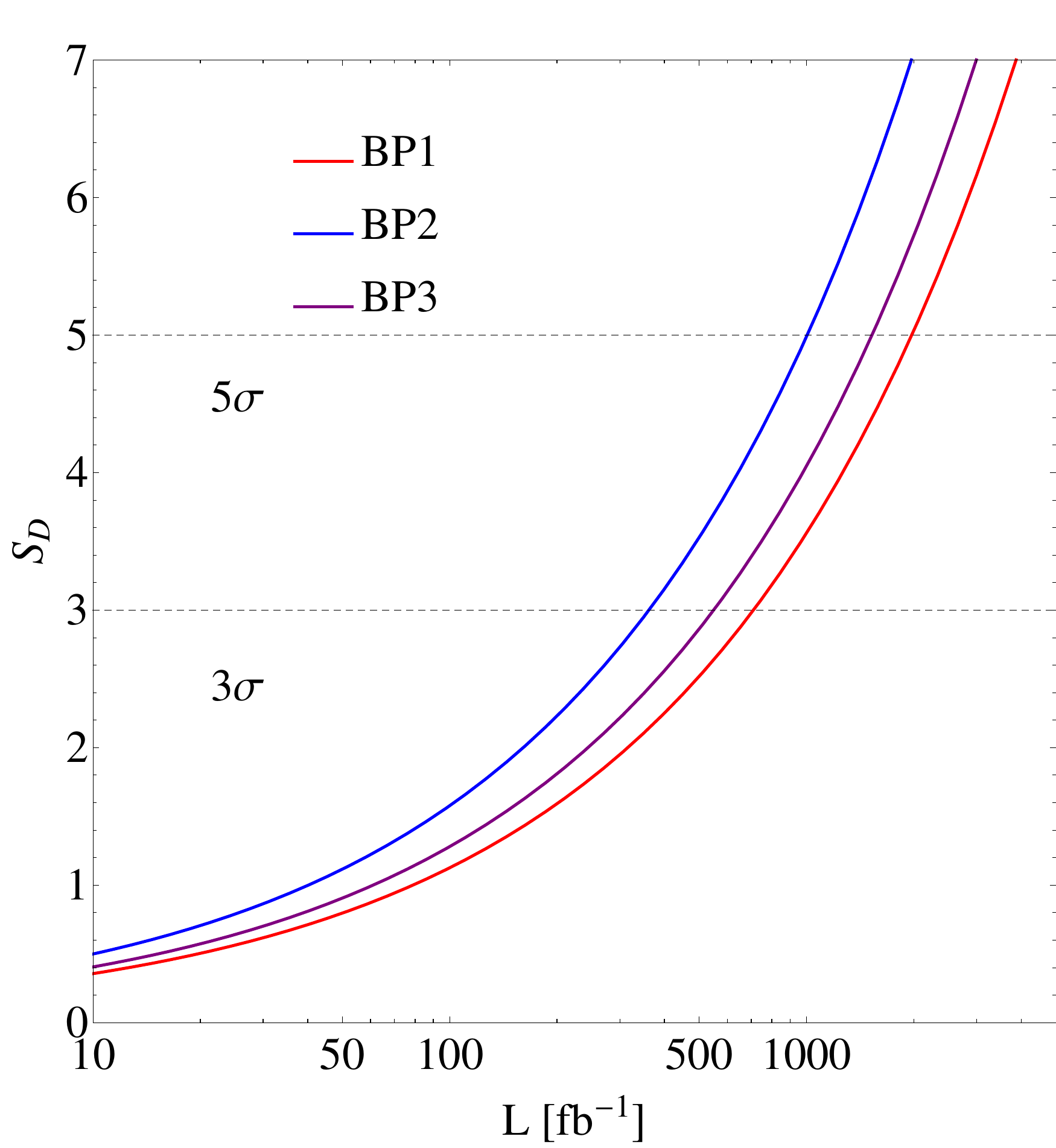}
\hspace{5mm}
\includegraphics[width=0.45\textwidth]{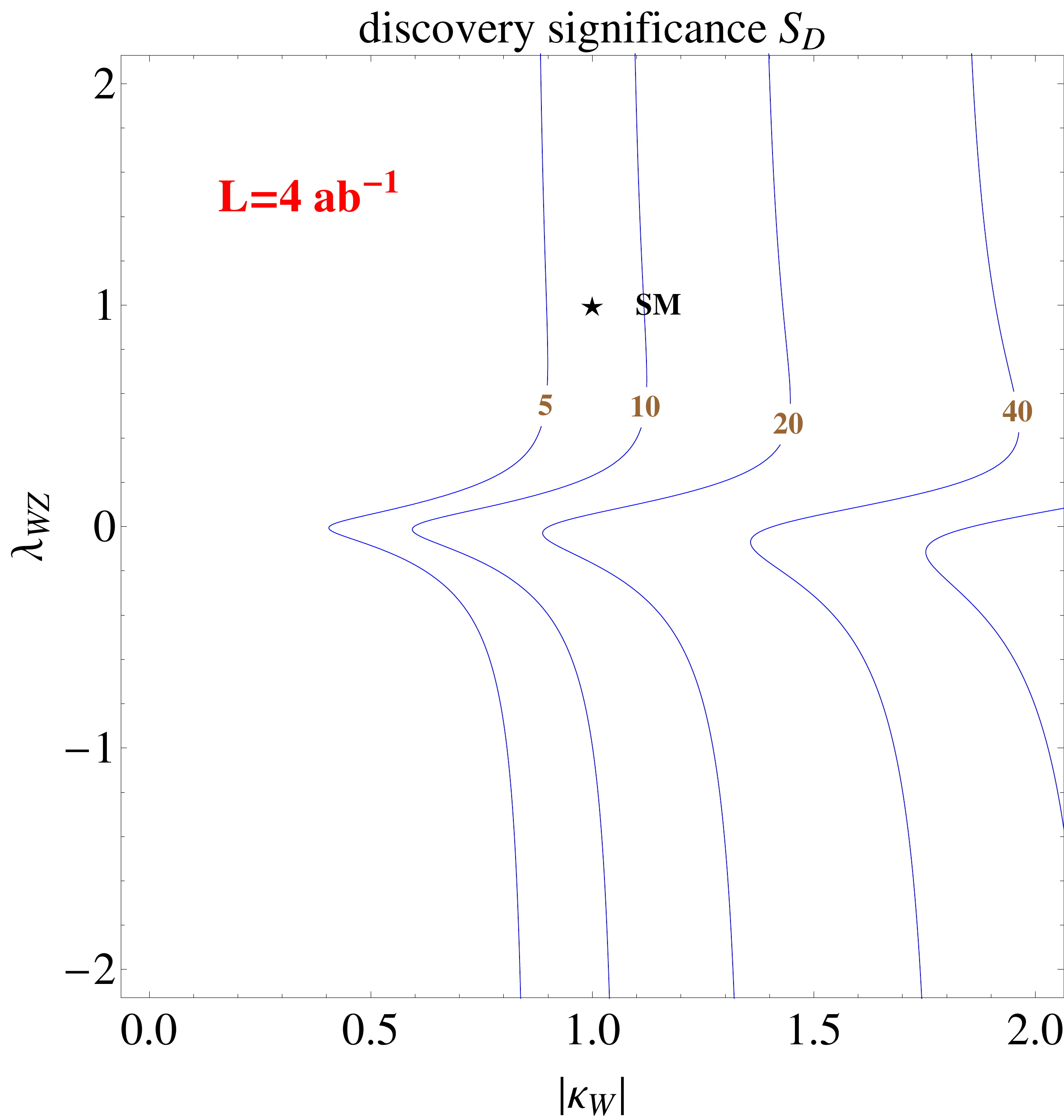}
\caption{Discovery potential of $e^+e^-\to W^+W^-H$ in the $\ell^{\pm}\ell^{\pm}\ell^{\mp}jj$ channel for scheme A at the 500-GeV ILC with the beam polarization $P(e^-,e^+)=(-0.8,+0.3)$.  Left: signal significance of the three scenarios as a function of the integrated luminosity.  Right: contours of signal significance in the $|\kappa_W|$-$\lambda_{WZ}$ plane, assuming an integrated luminosity of $4\abi$.}
\label{fig:discovery_hww_A}
\end{figure}

For scheme A, where the couplings of $H$ to the SM fermions are the same as those of the SM Higgs boson, the discovery potential of \eewwh in the \llljj channel at the 500-GeV ILC with the beam polarization $P(e^-,e^+)=(-0.8,+0.3)$ is shown in Fig.~\ref{fig:discovery_hww_A}, which is worse than the $jj\ell^{\pm}bb$ channel.  The left plot shows that the required integrated luminosities for a $5\sigma$ discovery through this channel are about $2\abi$, $1\abi$ and $1.5\abi$ for scenarios (BP1, BP2, BP3), respectively.  The right plot shows that a $5\sigma$ signal can be observed with $L = 4\abi$ if $|\kappa_W|\gtrsim 0.9$, irrespective of the value of $\lambda_{WZ}$. Similar to the \jjlbb channel, the \llljj channel is also more sensitive to negative $\lambda_{WZ}$ as compared to positive $\lambda_{WZ}$.  Again, the $\lambda_{WZ}\alt 0.46$ region has a better sensitivity, and the asymmetric shape in the contours enables us to solve the $\pm \lambda_{WZ}$ ambiguity. However, since the discovery potential of the $\ell^{\pm}\ell^{\pm}\ell^{\mp}jj$ is partially determined by $\lambda_{WZ}$ as ${\cal B}(H \to WW) \propto \kappa_W^2$, it is impossible to have a $5\sigma$ signal if $|\kappa_W|\lesssim 0.4$, irrespective of the value of $\lambda_{WZ}$.

\begin{figure}[!htb]
\centering
\includegraphics[width=0.45\textwidth]{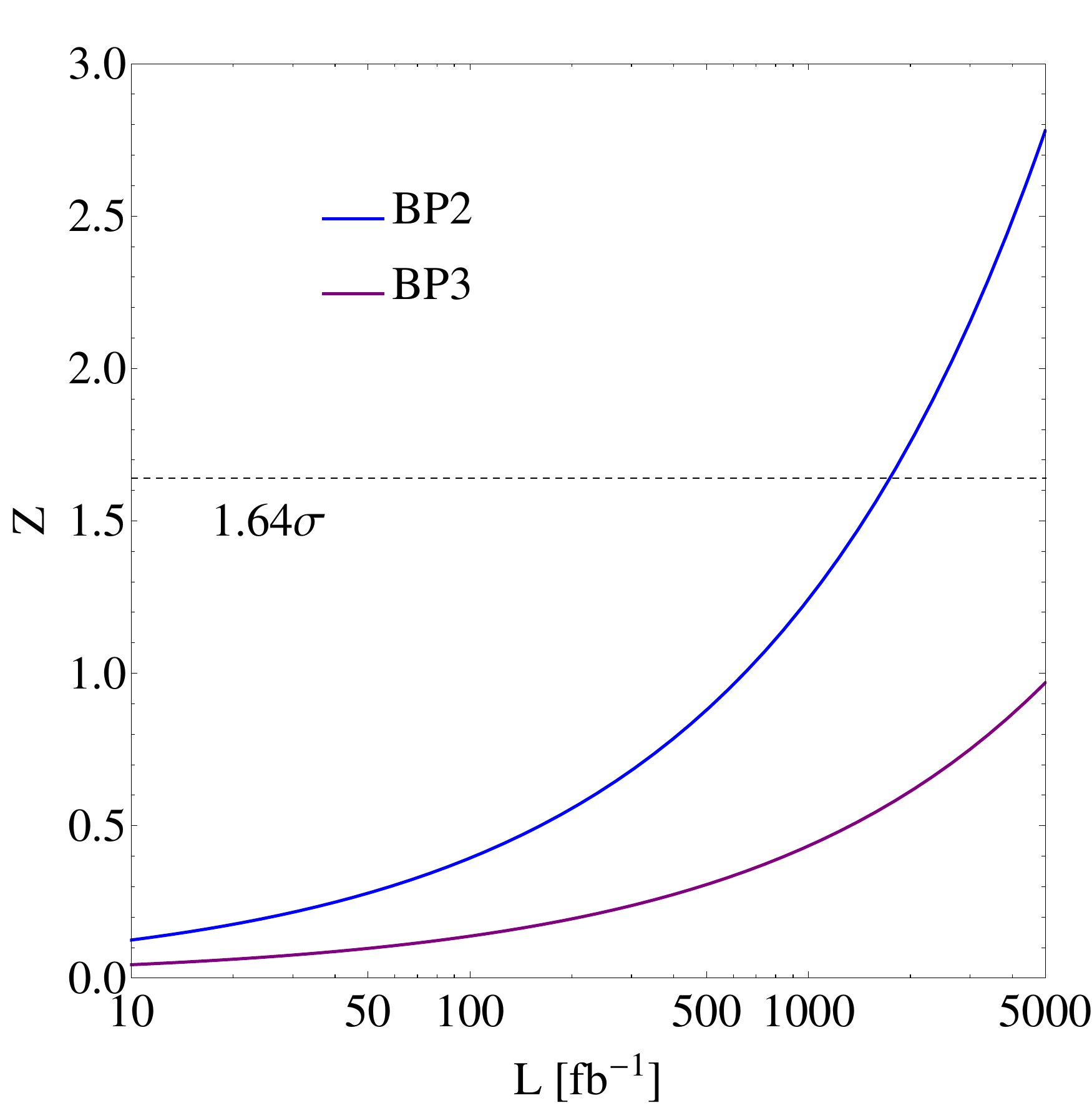}
\hspace{5mm}
\includegraphics[width=0.45\textwidth]{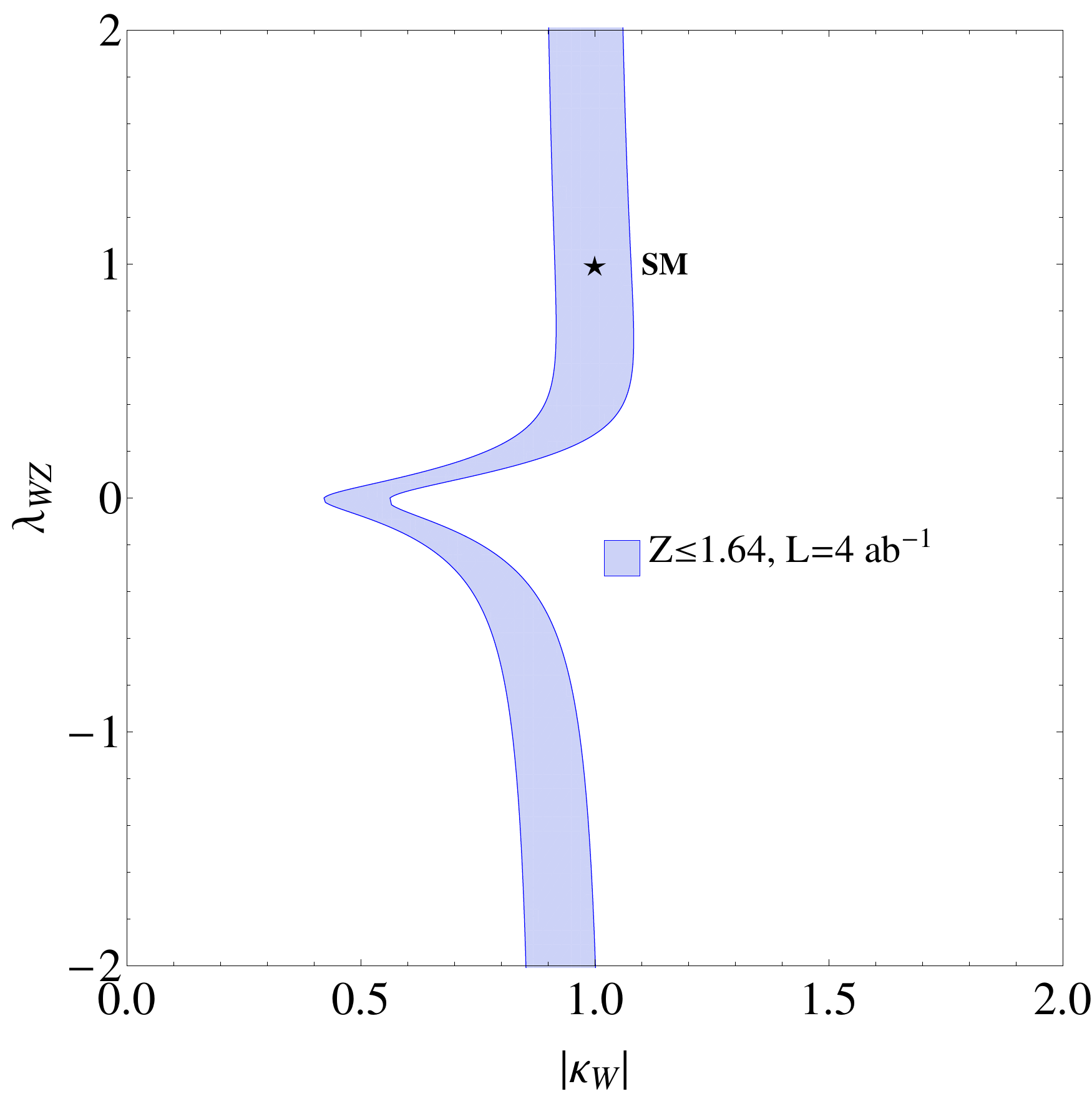}
\caption{Power to discriminate $\lambda_{WZ}$ in the $\ell^{\pm}\ell^{\pm}\ell^{\mp}jj$ channel for scheme A at the 500-GeV ILC with the beam polarization $P(e^-,e^+)=(-0.8,+0.3)$.  Left: discriminating the (BP2, BP3) scenarios from the SM scenario (BP1).  Right: region in the $|\kappa_W|$-$\lambda_{WZ}$ plane that satisfies $Z \le 1.64$, assuming an integrated luminosity of $4\abi$.}
\label{fig:exclusion_hww_A}
\end{figure}

In Fig.~\ref{fig:exclusion_hww_A}, we show the ability to discriminate $\lambda_{WZ}$.  The left plot shows that the BP2 scenario $(\lambda_{WZ}=-1)$ can be distinguished from the SM (BP1) at 95\% C.L. with an integrated luminosity of about $1.74\abi$.  The BP3 scenario, however, requires a much higher luminosity. As shown in the the right plot, if $|\kappa_W|> 1$ the negative $\lambda_{WZ}$ region in the plotted area can be completely excluded at 95\% C.L. through the \eewwh process in the \llljj channel. 

\begin{figure}[!htb]
\centering
\includegraphics[width=0.5\textwidth]{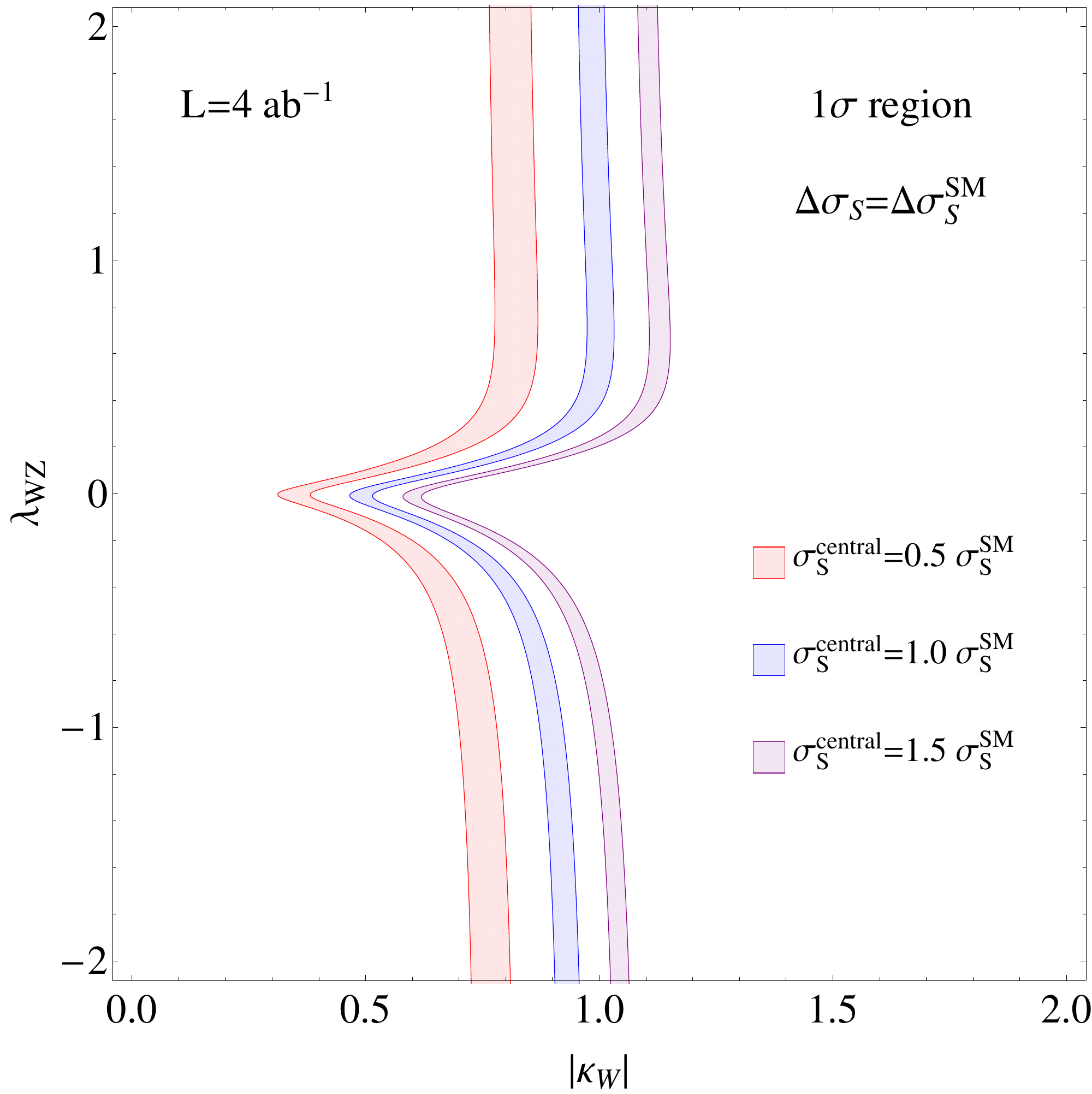}
\caption{$1\sigma$ regions in the $|\kappa_W|$-$\lambda_{WZ}$ plane when the measured signal cross section has the central value $\sigma_S^{\text{central}}= (0.5, 1, 1.5) \times \sigma_S^{\text{SM}}$ and an error bar assumed to be the same as the SM expectation, assuming the 500~GeV ILC with the beam polarization $P(e^-,e^+)=(-0.8,+0.3)$ and the integrated luminosity of $4\abi$.}
\label{fig:precision_hww_A}
\end{figure}

Fig.~\ref{fig:precision_hww_A} plots the $1\sigma$ region allowed by the assumed measurements given in Eq.~\eqref{eq:central_values}.  As compared to the $jj\ell^{\pm} bb$ channel, the allowed regions in the $\ell^{\pm}\ell^{\pm}\ell^{\mp}jj$ channel are slightly more symmetric with respect to $\lambda_{WZ}=0$ and thus less sensitive to the sign of $\lambda_{WZ}$. 

We can also obtain the relative error on $\lambda_{WZ}$ by combining the \eewwh process in the \llljj channel and the measurement of $|\kappa_Z|$ at the HL-LHC, which gives $\delta\lambda_{WZ}/\lambda_{WZ}<0.071$.

\subsubsection{Scheme B}

\begin{figure}[t!hb]
\centering
\includegraphics[width=0.45\textwidth]{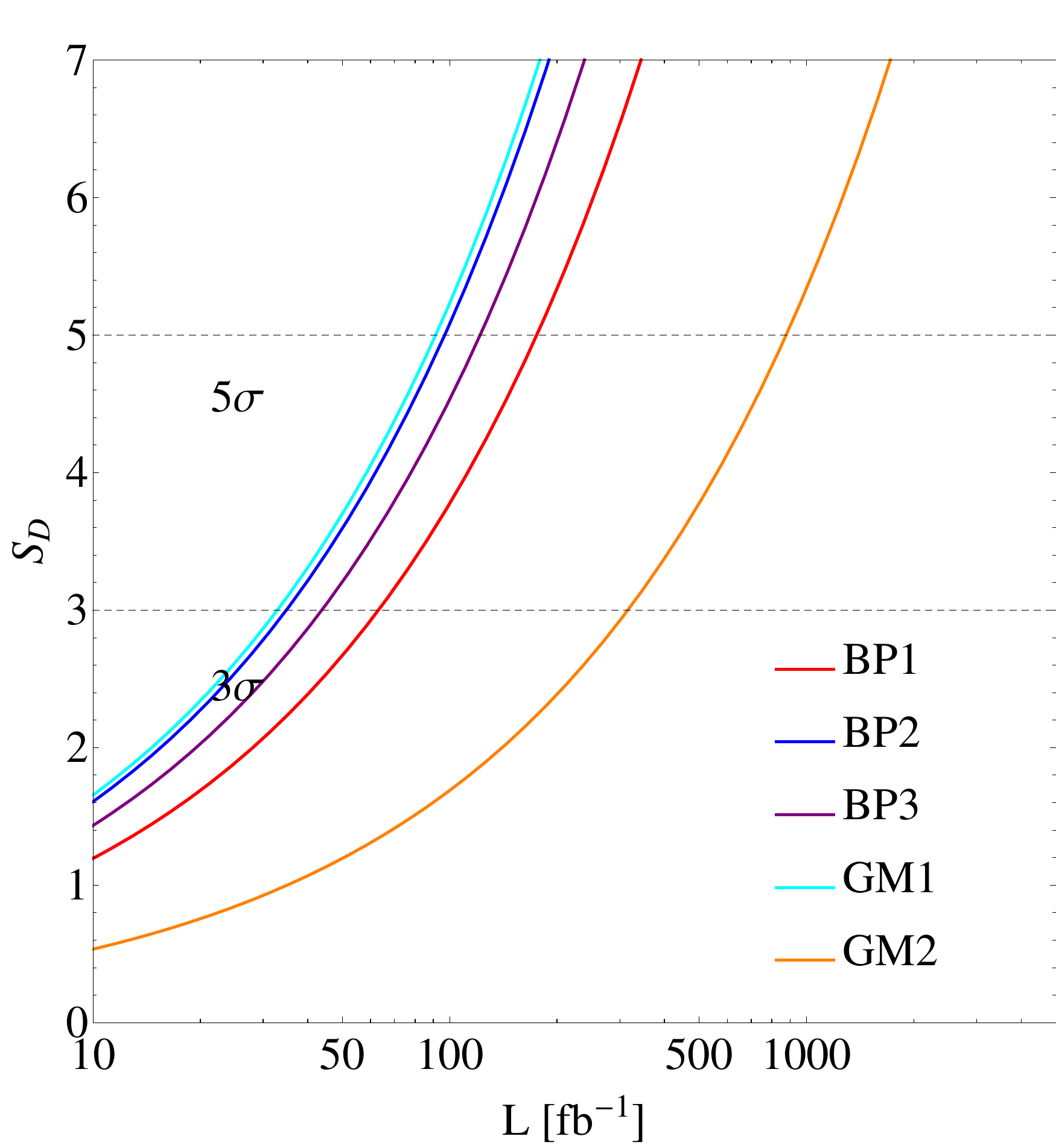}
\hspace{5mm}
\includegraphics[width=0.45\textwidth]{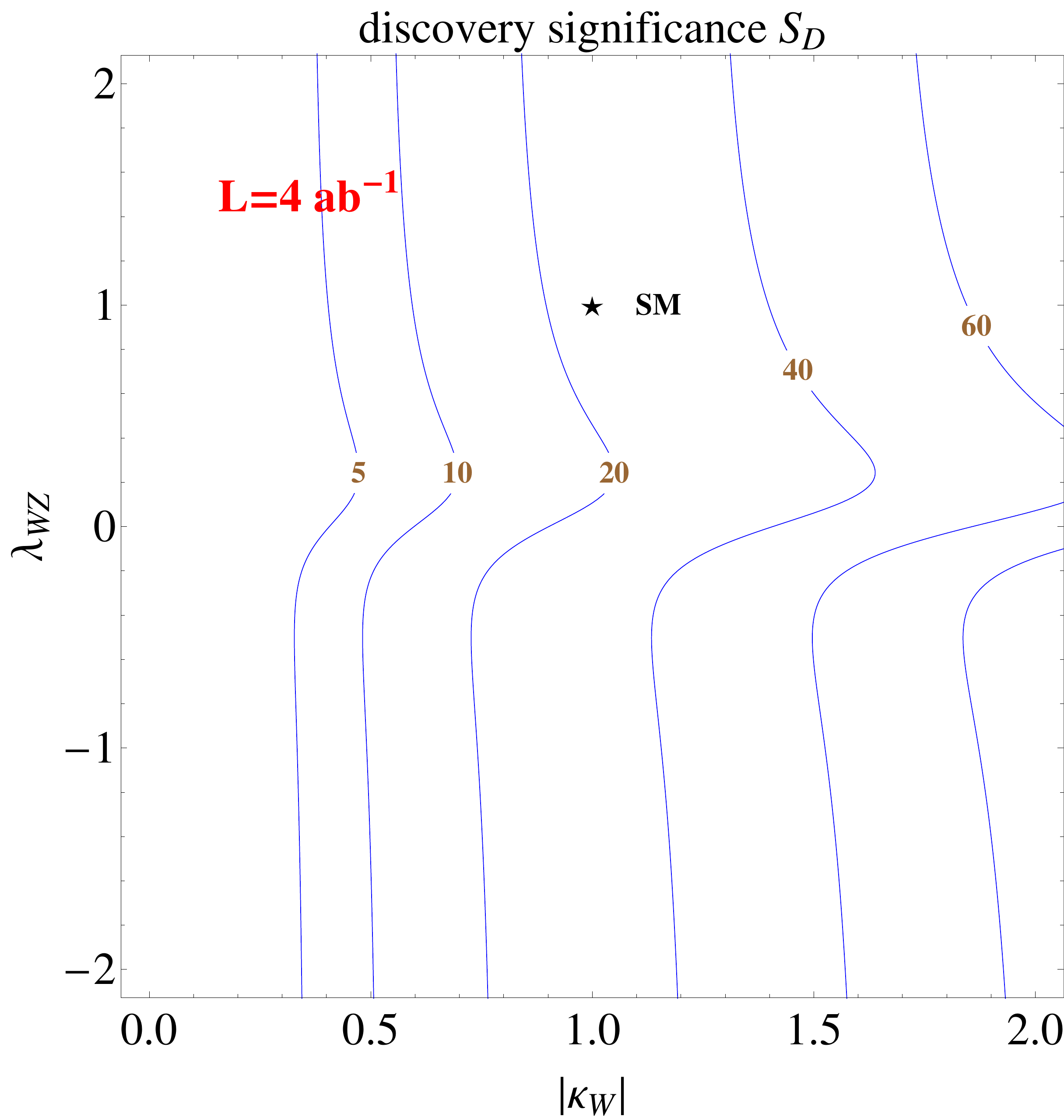}
\caption{Discovery potential of $e^+e^-\to W^+W^-H$ in the $\ell^{\pm}\ell^{\pm}\ell^{\mp}jj$ channel for scheme B at the 500-GeV ILC with the beam polarization $P(e^-,e^+)=(-0.8,+0.3)$.  Left: signal significance of the five scenarios (defined in the main text) as a function of the integrated luminosity.  Right: contours of signal significance in the $|\kappa_W|$-$\lambda_{WZ}$ plane, assuming an integrated luminosity of $4\abi$.}
\label{fig:discovery_hww_B}
\end{figure}

For scheme B where $H$ is fermiophobic, the \llljj channel provides a unique signature.  Such a scheme happens to the fiveplet Higgs boson $H_5^0$ in the GM model~\cite{Georgi:1985nv,Chanowitz:1985ug,Chiang:2012cn,Chiang:2015kka,Chiang:2015rva}, where $\lambda_{WZ} = -1/2$.  For this, we introduce two more benchmark scenarios:
\begin{align}
\begin{split}
&\text{GM1:}\quad |\kappa_W|=1 ~,\ \lambda_{WZ}=-\dfrac{1}{2} 
~, \\
&\text{GM2:}\quad |\kappa_W|=\dfrac{1}{2} ~,\ \lambda_{WZ}=-\dfrac{1}{2}
~.
\end{split}
\end{align}
For concreteness, we keep $m_H = 125$~GeV in the following numerical analysis, noting that it is conceptually analogous to apply the method on an exotic Higgs boson of a different mass.

Fig.~\ref{fig:discovery_hww_B} shows the discovery potential of the \eewwh process in the $\ell^{\pm}\ell^{\pm}\ell^{\mp}jj$ channel for scheme B with five benchmark scenarios of the couplings (BP1, BP2, BP3, GM1, GM2) assuming that $H$ is fermiophobic. Compared to scheme A, the sensitivity is notably improved due to an increase in the branching ratio of $H\to W^+W^-$.  According to the left plot, $\lambda_{WZ} = -1/2$ can be confirmed with an integrated luminosity of about $90\fbi$ and $870\fbi$ for GM1 and GM2, respectively. Note that the sensitivities of BP2 and GM1 are close to each other since for $|\kappa_W|=1$ a change of $\lambda_{WZ}$ from $-1$ to $-1/2$ can increase the production cross section and reduce the decay branching ratio at the same time. The right plot shows that the \eewwh process can be discovered with an integrated luminosity of $4\abi$ provided $|\kappa_W|\gtrsim 0.35$.

\begin{figure}[t!hb]
\centering
\includegraphics[width=0.45\textwidth]{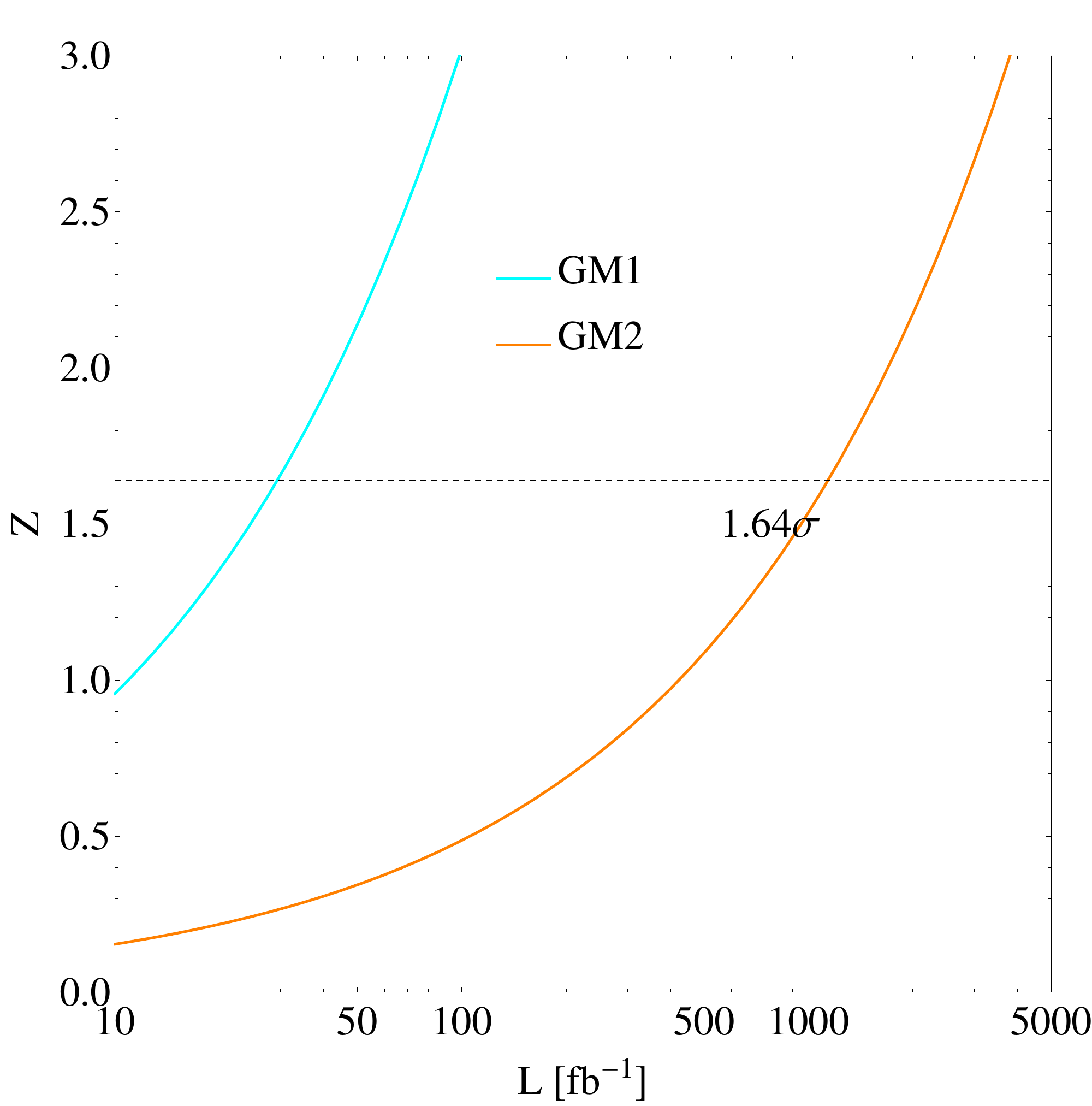}
\hspace{5mm}
\includegraphics[width=0.45\textwidth]{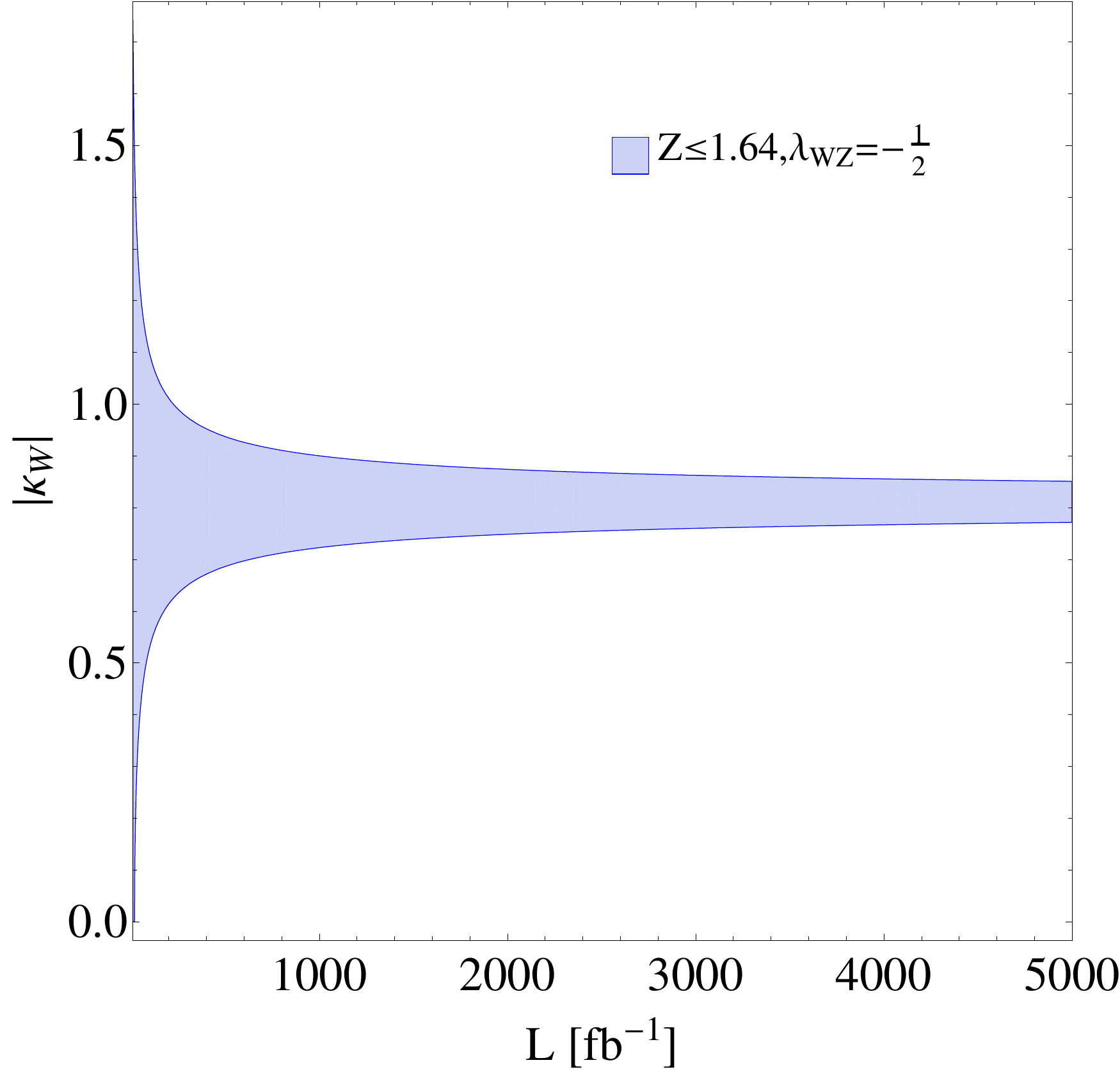}
\caption{Power to discriminate $\lambda_{WZ} = -1/2$ from the SM value in the $\ell^{\pm}\ell^{\pm}\ell^{\mp}jj$ channel for scheme B at the 500-GeV ILC with the beam polarization $P(e^-,e^+)=(-0.8,+0.3)$.  Left: discriminating the GM1 and GM2 scenarios from the SM.  Right: the assumption of $\lambda_{WZ}=-1/2$ can be distinguished from the SM assumption of $\lambda_{WZ}=1$ at 95\% C.L. for a varying $|\kappa_W|$ out of the blue region.}
\label{fig:exclusion_hww_B}
\end{figure}

For this scheme, we are mostly interested in discriminating $\lambda_{WZ} = -1/2$ from the SM value of $\lambda_{WZ}=1$. In Fig.~\ref{fig:exclusion_hww_B}, we show the discriminating power for the two GM scenarios from the SM Higgs boson. From the left plot, we find that $H$ in the GM1 and GM2 scenarios can be distinguished from the SM case with an integrated luminosity of about $30\fbi$ and $1.14\abi$, respectively. As a comparison, it is claimed that the $\lambda_{WZ}=1$ and $-1/2$ cases can be discriminated at 95\% C.L. at the LHC through $H\to ZZ^{*}\to 4\ell$ with an integrated luminosity of $2\abi$~\cite{Chen:2016ofc}. The right plot shows the $Z\leq 1.64\sigma$ region within which the $\lambda_{WZ}=-1/2$ case cannot be excluded at 95\% C.L.

\section{Summary and Conclusions}
\label{sec:summary}

In examining properties of a Higgs boson, the discovered 125-GeV SM-like and any new Higgs bosons alike, it is important to determine separately its couplings to the $W$ and $Z$ bosons, including their relative sign and the magnitudes of these couplings.   In this work, we have proposed to use the $W^+W^-H$ production process to determine the sign and magnitude of the ratio of the $HWW$ and $HZZ$ couplings at future colliders.  This process includes Feynman diagrams involving both couplings at tree level.  We found that the sensitivity of this process at the LHC  is low.
We therefore focused our study on the \eewwh process at a future $e^+e^-$ collider.  For concreteness, we took the 125-GeV Higgs boson as an explicitly example.  In this case, we have found that a 500-GeV $e^+e^-$ collider with appropriately polarized beams is suitable for such an analysis.

We have considered two decay channels of the Higgs boson and thus divided our discussions according to two kinds of final states \jjlbb and \llljj.  In the latter case, we further examined the cases when $H$ is SM-like or fermiophobic in its couplings to the SM fermions.  We analyzed a few benchmark coupling scenarios in terms of the scale factors $\kappa_W$ and $\kappa_Z$.  For all the scenarios considered, we obtained the discovery potential for determining the ratio of $HWW$ and $HZZ$ couplings, $\lambda_{WZ}$.  The results were shown in the plane of $|\kappa_W|$ and $\lambda_{WZ}$.  Due to the destructive interference between the diagrams involving the $HWW$ coupling and that involving the $HZZ$ coupling, we found that the \eewwh process had different sensitivities for different values of $\lambda_{WZ}$.  In particular, the negative $\lambda_{WZ}$ scenarios generally require less luminosity to reach a $5\sigma$ discovery.

To discriminate a particular $\lambda_{WZ}$ scenario from the SM case, we further made use of the log likelihood ratio.  By evaluating such a quantity, we have obtained the required integrated luminosities to discriminate two theory assumptions at 95\% C.L. as well as the region in the plane of $|\kappa_W|$ and $\lambda_{WZ}$ that satisfied $Z \le 1.64$ with an integrated luminosity of $4\abi$.  We have also investigated the impact of three possible signal cross section outcomes and plotted the $1\sigma$ region in the $|\kappa_W|$-$\lambda_{WZ}$ plane.  It was found that for a smaller (larger) cross section, the \eewwh process is less (more) sensitive to the sign of $\lambda_{WZ}$.  By combining the cross section measurement of the proposed process and the measurement of $|\kappa_Z|$ at the HL-LHC, it will be straightforward to determine $\lambda_{WZ}$ at a high precision, as compared to purely from the measurements of $|\kappa_W|$ and $|\kappa_Z|$ at the HL-LHC.


Finally, we would like to emphasize that our method of determining $\lambda_{WZ}$ for the 125-GeV Higgs boson can be readily applied to another new Higgs boson that couples to both $W$ and $Z$ bosons.  We also note that a similar tree-level interference effect in the $pp\to jjW^{\pm}H$ and $pp\to jjZH$ processes can be used to determine $\lambda_{WZ}$, the result of which will be given in a separate work~\cite{inpreparation}.

\acknowledgments
We would like to thank Matti Heikinheimo for communications regarding details of Ref.~\cite{Gabrielli:2013era}, and Giovanna Cottin, Jinmian Li, Tanmoy Modak, Sichun Sun, Yi-Lei Tang and Bin Yan for valuable discussions. 
This work was supported  in part by Ministry of Science and Technology under Grant No.~104-2628-M-002-014-MY4, No.~104-2112-M-002-015-MY3 and No.~106-2112-M-002-003-MY3, and National Natural Science Foundation of China under Grant No.~11575110, 11575111, 11655002, 11735010,  Natural  Science Foundation of Shanghai under Grant  No.~15DZ2272100. 

\bibliographystyle{apsrev}
\bibliography{reference}
%
\end{document}